\documentclass[prl,aps,epsf,twocolumn,superscriptaddress,longbibliography]{revtex4-2}
\usepackage{natbib}
\usepackage{amsmath,amssymb,amsfonts,bbm,float,graphics,epsfig,epstopdf,color,verbatim,tabularx,bm,multirow,hyperref,ulem,times,subfigure}
\hypersetup{
  colorlinks, linkcolor={blue},
  citecolor={blue}, urlcolor={blue}
}
\usepackage{textcomp}
\usepackage{times}
\usepackage{tabularx}
\usepackage{graphicx}
\usepackage{float}
\usepackage{latexsym,amsmath,amssymb,bm,euscript}
\usepackage{color}
\usepackage{subfigure}
\usepackage{epstopdf}

\usepackage{float}
\usepackage{soul}

\usepackage{hyperref}
\usepackage{bookmark}
\usepackage{CJK}
\usepackage{graphicx}
\usepackage[english]{babel}
\usepackage{amsmath}
\usepackage{multirow}
\usepackage{xcolor}

\newcommand{\updated}[1]{#1}

\begin{document}

\title{Spectra of Magnetoroton and Chiral Graviton Modes of Fractional Chern Insulator}

\author{Min Long}
\affiliation{Department of Physics and  HK Institute of Quantum Science \& Technology, The University of Hong Kong, Pokfulam Road,  Hong Kong SAR, China}

\author{Hongyu Lu}
\affiliation{Department of Physics and  HK Institute of Quantum Science \& Technology, The University of Hong Kong, Pokfulam Road,  Hong Kong SAR, China}

\author{Han-Qing Wu}
\affiliation{Guangdong Provincial Key Laboratory of Magnetoelectric Physics and Devices,
School of Physics, Sun Yat-sen University, Guangzhou 510275, China}

\author{Zi Yang Meng}
\email{zymeng@hku.hk}
\affiliation{Department of Physics and  HK Institute of Quantum Science \& Technology, The University of Hong Kong, Pokfulam Road,  Hong Kong SAR, China}

\date{\today}

\begin{abstract}
Employing the state-of-the-art time-dependent variational principle (TDVP) algorithm, we compute the spectra of charge-neutral excitations in the $\nu=1/2$ (bosonic) \updated{ and $1/3$ (fermionic) fractional Chern insulator (FCI)} on the Haldane honeycomb lattice model. The magnetoroton visualized from the dynamic density structure factor acquires a minimum gap at finite momentum that can go soft with increasing interaction and give rise to a charge density wave (CDW) at the same wavevector. As the system approaches the FCI-to-CDW transition point, we observe a pronounced sharpening of the roton mode, \updated{suggesting that the magnetoroton behaves more like a quasiparticle as it softens}. 
Notably, this occurs while the single-particle gap remains finite.
Besides the magnetoroton at finite momentum, we also construct quadrupolar chiral operators in a discrete lattice and resolve the chiral graviton mode around the $\Gamma$ point of the Brillouin zone. Furthermore, we show the different chiralities of the gravitons of FCIs with opposite-sign Hall conductance for the first time.
\updated{Our results offer the clear spectral observations of magnetoroton and chiral graviton in FCI lattice models and will have relevance towards systematic understanding of elementary excitation in FCIs}.
\end{abstract}

\maketitle

\noindent{\textcolor{blue}{\it Introduction.}---}
\updated{Charge-neutral excitation in fractional quantum Hall (FQH) states has drawn immense attention in the past decades. The pioneer work by Girvin, MacDonald, and Platzman (GMP) established the magnetoroton theory of the Laughlin wave function based on the single-mode approximation (SMA)~\cite{Girvin_1985}, which interpreted the charge-neutral mode as a collective density excitation~\cite{scarola2000rotons,Du1993Experimental,Pinczuk1993Observation,Kang2001Observation,Igor2009Dispersion}
whose softening triggers a phase transition from FQH  to CDW~\cite{Girvin_1985,Kumar_2022,tsui2024direct,song2024intertwined}. Later on, Haldane propose a geometric description of charge neutral mode on long wavelength limit, reflecting oscillation of intrinsic metric of Landau orbitals~\cite{Haldane2011}, known as chiral graviton mode (CGM) with spin angular momentum $2$ ~\cite{Qiu2012,Liou2019,Johri2016,Haldane2021,Liou2019Chiral, Wang2022,Yang_2012}. CGM has recently been observed in the experiment using circularly polarized resonant inelastic light scattering~\cite{Liang_2024}. While the magnetoroton mode tends to break translational symmetry and trigger FCI-CDW transition, the graviton mode could break rotational symmetry and is responsible for anisotropic transport in the FQH nematics ~\cite{Xia_2011,Yang_2020,Pu_2024,Regnault_2017} and describes nonequilibrium dynamics after magnetic quench~\cite{Liu2018Geometric,Balram2022Very}.}


\updated{FCI is the lattice analog of FQH~\cite{Parameswaran_2013,Sheng2011,Wang2011,Regnault2011} and could exist without external magnetic field due to the orbital magnetism~\cite{Sheng2011,neupert2011fractional,Tang2011}m, also known as the fractional quantum anomalous Hall (FQAH) states. FQAH was recently observed in  MoTe$_2$ bilayers~\cite{caiSignature2023,zengThermodynamic2023,xu2023_fci, park2023_fqah} and in rhombohedral pentalayer graphene/hBN moir\'e superlattices~\cite{ZLu2023fqh_graphene}.
In parallel with FQH, the study of charge-neutral excitations in FCI mostly adopts GMP ansatz, where the Hilbert space is restricted to a single band and perform exact diagonalization (ED),  yielding similar behavior of charge neutral excitation as FQH~\cite{Haldane1985,Haldane2021,Girvin_1985,Kumar_2022,Liou2019,Wang2022,Yang_2020,Yang_2012,Repellin_2014,Liu2024,Kousa2025Theory}. However, the investigation of FCI present several unique aspects. The energy scales in the FCI system are more complex, where the interaction strength could be larger than the band gap, making the band mixing effect non-negligible~\cite{yu2024moire,Abouelkomsan2024Band,ZLu2023fqh_graphene}. Moreover, the importance of spectral weight remains unrevealed, as it contains information on quasi-particle lifetime and is directly related to full width at half maximum analysis in the Raman scattering experiment~\cite{Liang_2024}. Furthermore, investigations of CGMs have predominantly employed momentum-space approaches, while the detection of CGM based on a real-space approach (considering the lattice effect) remains largely unexplored. 
Given the situation, a comprehensive study of the spectral properties of magnetoroton and CGM in FCI will be of great importance to guide future experiments and provide unbiased verifications for the development of unifying theories for FQH and FCI systems.}

\begin{figure*}[htp!]
\centering
\includegraphics[width=0.9\textwidth]{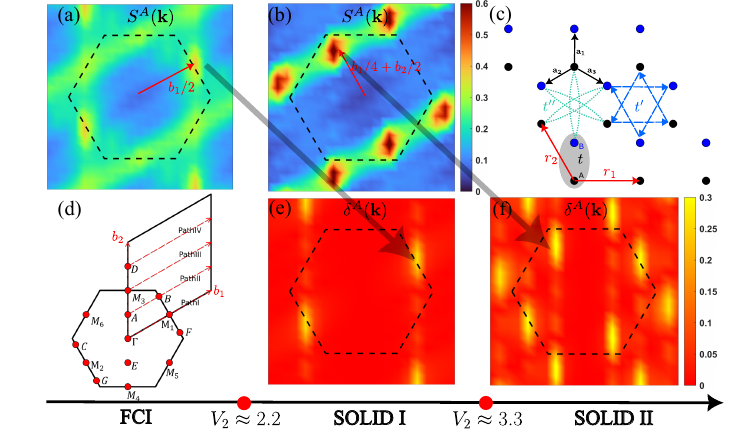}
\caption{\textbf{Static density structure factor $S^A(\mathbf{k})$ and charge order parameter $\delta^A(\mathbf{k})$ for $1/2$ bosonic FCI}. Along the parameter path we choose, the system undergoes a sequence of FCI-Solid I-Solid II phase transitions. (a), (b) are the $S^A(\mathbf{k})$ of FCI phase ($V_1 = 4, V_2 = 2$) and Solid I phase ($V_1 = 4, V_2 = 3$). (c), (d) are the real and reciprocal spaces of the honeycomb lattice FCI model in Eq.~\eqref{eq:eq1}. In (c), $\mathbf{r_1} = (\sqrt{3},0), \mathbf{r_2} = (-\sqrt{3}/2,3/2)$ are the primitive lattice vectors. The NN, NNN, and NNNN hoppings are depicted with black, blue and green colours, respectively. In (d), the solid hexagon is the first Brillouin zone (BZ) of the model, with high symmetry points labelled. The four paths that are discrete in $\mathbf{b_2}$ direction with step $\mathbf{b_2}/4$ are where we compute the spectra under the cylinder geometry in DMRG calculation. (e), (f) are the $\delta^A(\mathbf{k})$ of Solid I phase ($V_1 = 4, V_2 = 3$) and Solid II phase ($V_1 = 4, V_2 = 5$). The roton mode locates at $\mathbf{b_1}/2$ in (a) condensates and leads to charge order in (e), while similar $\mathbf{b_1}/4+\mathbf{b_2}/2$ roton in (b) condenses and give rise to the charge order in (f). The data that determine of the transition points are given in Supplemental Material (SM)~\cite{suppl}.}
        \label{fig:fig1}
    \end{figure*}

In this letter, we fill in such a knowledge gap by studying the spectral properties of FCI realised by hard-core bosons and spinless fermions on the Haldane honeycomb lattice model when the flat band is partially filled~\cite{Wang2011,Sheng2011}. And the $1/2$ bosonic FCI therein can transit to two CDW phases as a function of the repulsive density-density interaction~\cite{luVestigial2024}. We employ the state-of-the-art time-dependent variation principle (TDVP)~\cite{Haegeman2011Time,Haegeman2016Unifying} within large-scale density matrix renormalization group (DMRG) simulaltion~\cite{White1992Density} to reveal the spectroscopic signatures for the FCI state and its transition to CDW orders. We find that (1) the charge-neutral magnetoroton can be visualized from the dynamic density structure factor. The magnetoroton mode could go soft and finally leads to the CDW order at the same momentum of the roton minimum, whereas the single-particle gap remains large and intact throughout; (2) the spectral weight of roton become sharper and highly concentrated around its minimum when approaching the phase boundary, manifesting the roton condensation character of the FCI-CDW transition; (3) inside the $1/2$ bosonic FCI, a strong gapped chiral graviton mode -- probed by quadrupolar density correlation function -- with angular momentum-2 is clearly present and its chirality can be tuned by flipping the Chern number $\pm1$ of the flat band; \updated{(4) the chiral signal observed in the $1/3$ fermionic FCI exhibits significantly reduced spectral weight compared to the bosonic case, suggesting potentially distinct behavior of CGM in fermionic FCI.}  Our results show, in an unbiased manner, the robustness and tunability of the charge-neutral excitations in the FCI states and provide guidance towards their spectroscopic detection in experiments. 

\noindent{\textcolor{blue}{\it Model and Method.}---} We consider the topological flat band Hamiltonian on honeycomb lattice~\cite{Wang2011,Sheng2011,Grushin2015Characterization,luVestigial2024,Luo_2020} \updated{(in the analysis of magnetoroton and charge excitation in the text, we focus on the $1/2$ bosonic FCI, which could be realized in cold atom system~\cite{L_onard_2023}, leaving the $1/3$ fermionic FCI results in the supplemental material(SM)~\cite{suppl}. While for CGM in the main text we display both $1/2$ bosonic FCI and $1/3$ fermionic FCI results)}:
\begin{equation}
    \begin{aligned}
        H&=  -t\sum_{\langle i, j\rangle} \left(b_i^{\dagger} b_j e^{i\phi_{ij}}+
        \text { H.c. }\right)- t^{\prime}\sum_{\langle\langle i, j\rangle\rangle}\left(b_i^{\dagger} b_j+
        \text { H.c. }\right) \\
        & -t^{\prime \prime}\sum_{\langle\langle\langle i, j\rangle\rangle\rangle} \left(b_i^{\dagger} b_j+\text { H.c. }\right) + V_1 \sum_{\langle i, j\rangle} n_i n_j+V_2 \sum_{\langle\langle i, j\rangle\rangle} n_i n_j,
    \end{aligned}
    \label{eq:eq1}
\end{equation}
where $b_i^{\dagger}$ creats a hard core boson at site $i$, $\langle ... \rangle,\langle 
\langle ...\rangle\rangle$, 
and $\langle\langle ...\rangle\rangle\rangle$ denote the nearest neighbour (NN), next nearest neighbour (NNN) and next next nearest neighbour (NNNN) sites, respectively, as shown in Fig.~\ref{fig:fig1} (c). In the study of $1/3$ fermionic FCI, we replace $b_i^{\dagger}$ and $b_i$ with  $c_i^{\dagger}$ and $c_i$, which stand for the creation and annihilation operators for spinless fermions \cite{Grushin2015Characterization}. 

To obtain the flat band with Chern number $C=1$, we set $t = 1$ (energy unit throughout the paper), $t^{\prime} = 0.6$ , $\phi = \pm 0.4\pi$ and $t^{\prime \prime} = -0.58$, following the literature~\cite{Wang2011,Sheng2011,luVestigial2024}. $V_1(V_2)$ refers to the amplitude of NN (NNN) repulsive interactions. We note that even without the neighboring interaction terms in Eq.~\eqref{eq:eq1}, the model still hosts a $1/2$ bosonic FCI when the flat band is half filled, which is attributed the the hard-core condition as an infinite on-site repulsion $V_0$. 


In our simulation, we consider a finite cylinder with $N_1\times N_2 = 4 \times16$ unit cells. The total number of sites is $N =2 N_1  N_2 $ corresponding bosons particle number of $1/2$ fills the flat band is $N_b = N_1  N_2 /2 = 32 $. For $1/3$ fermionic FCI, we consider a $N_1\times N_2 = 3 \times18$ cylinder. The details of the cylinder geometry and simulations are described in SM~\cite{suppl}.

\begin{figure}[h!]
    \centering
    \includegraphics[width=\columnwidth]{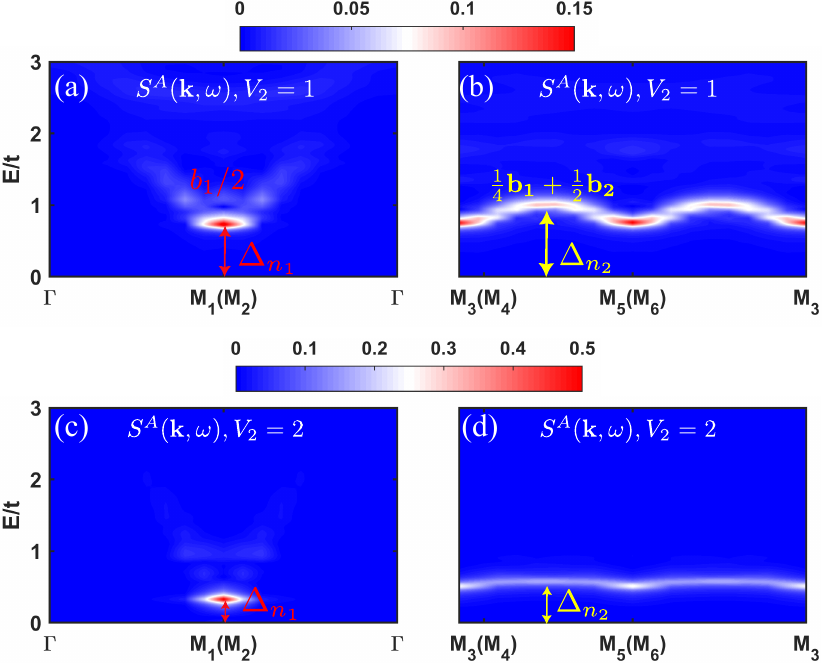}
    \caption{\textbf{Magnetoroton spectrum inside $1/2$ bosonic FCI phase.} (a), (b) are $S^A(\mathbf{k},\omega)$ deep inside FCI with $V_1 = 4, V_2 = 1$
    along Path I and Path III in Fig.~\ref{fig:fig1} (d). Two charge neutral rotons relevant with the CDW phases are marked with $\Delta_{n_1}$ and $\Delta_{n_2}$ with their gaps denoted by the red and yellow arrows, respectively.  (c) and (d) are the roton spectra near the FCI-Solid I phase boundary with $V_1 = 4, V_2 = 2$. The $\Delta_{n_1}$ is about to close with very sharp spectral weight and give rise to the Solid I phase with the charge order at same momentum. The $\Delta_{n_2}$ will persist into the Solid I phase and condense at the Solid I - Solid II transition (discussed in Fig.~\ref{fig:fig1} (b) and (f)).}
    \label{fig:fig2}
\end{figure}

Before discussing the spectra, we begin with the ground state phase diagram as shown in Fig.\ref{fig:fig1}, The system goes through a FCI-Solid I-Solid II phase transition triggered by the condensation of two successive rotons (with momenta $\mathbf{b_1}/2$ and $\mathbf{b_1}/4 + \mathbf{b_2}/2$ respectively) as we increase $V_2$ with fixed $V_1 = 4$. \updated{Such a successive phase transition is similar with a vestigial transition~\cite{Nie2017vestigial,Fernandes2019Vestigial,wangVesigial2021,sun2024vestigial,svistunov2015superfluid,grinenko2021state} with the translational symmetry of the lattice broken in a two-step manner~\cite{luVestigial2024}}. Here, static density structure factor $S^{A}(\mathbf{k}) = \frac{1}{N_A}  \sum_{ij} e^{-i\mathbf{k}\cdot(\mathbf{x_i-x_j})}\langle n_{i,A}n_{j,A}\rangle$ and charge order parameter $\delta^{A}(\mathbf{k}) = \frac{1}{\sqrt{N_A}}  \sum_{i} e^{-i\mathbf{k}\cdot\mathbf{x_i}} \langle n_{i,A}\rangle$ are measured to diagnosis the phases, where $A$ denotes the sub-lattice, $N_A$ denotes its site number, $i,j$ run over all unit cells. Since the results are the same for both sublattices, we mainly present the results on $A$ sublattice. The bright point at $\frac{\mathbf{b_1}}{2}$ in Fig.~\ref{fig:fig1} (a) denotes the location of the lowest roton minimal of FCI~\cite{Lu2024Thermodynamics,Lu2024Interaction,luVestigial2024}, the fact that the rotational symmetry is not recovered in the $S^{A}(\mathbf{k})$ is a consequence of the cylindrical geometry of the DMRG simulation. In this system there is a second roton in the FCI phase, and it is located at the $\mathbf{b_1}/4 + \mathbf{b_2}/2$ (it can be clearly visualised in the spectra in Fig.~\ref{fig:fig2} (b)), and once the lowest roton condenses and the system enters the Solid I phase, the measurement of $S^{(A)}(\mathbf{k})$ in Fig.~\ref{fig:fig1} (b) shows that the brightest density correlation is at this wave vector ($V_1=4, V_2=3$). The charge order of Solid I, however, can be seen from its order parameter $\delta^{A}(\mathbf{k})$ 
in Fig.~\ref{fig:fig1} (e) ($V_1=4, V_2=3$).  And the grey arrow connecting the momentum $\mathbf{b_1}/2$ from Fig.~\ref{fig:fig1} (a) to Fig.~\ref{fig:fig1} (e), symbolises the condensation of the roton and formation of the charge order. The process also happened in transition from Solid I to Solid II, symbolised by the grey arrow connecting the momentum $\mathbf{b_1}/4+\mathbf{b_2}/2$ of the second roton in Fig.~\ref{fig:fig1} (b) to the charge order in Fig.~\ref{fig:fig1} (f) ($V_1=4, V_2=5$). \updated{The real space charge occupation pattern of two Solid phases is shown in SM~\cite{suppl}}.

\noindent{\textcolor{blue}{\it Magnetorotons.}---}With the ground state phase diagram at hand, we now investigate the spectra of magnetoroton in our system. 
The dynamic density structure factor is defined as:
\begin{equation}
    \begin{aligned}
        S^A(\mathbf{k},\omega) &= \frac{1}{\sqrt{N_tN_A}}  \sum_{jl} \ e^{i(\omega+i\eta) t_l } e^{-i\mathbf{k}(\mathbf{x_j-x_0})}\big( \langle n_{j,A}(t_l)n_{0,A}\rangle \\ &
        -\langle n_{j,A}\rangle \langle n_{0,A}\rangle\big)
    \end{aligned}
\end{equation}
where $n_{j,A}(t_l) = e^{iHl\Delta t}n_{j,A}e^{-iHl\Delta t}$ as $l$ runs from $0$ to $N_t$ and we choose $\eta = 0.05$ to broaden the peaks. We time evolve $n_{0,A}|\psi_0\rangle$ with $|\psi_0\rangle$ the ground state wave function and $n_{0,A}$ defined in the bulk of the cylinder~\cite{suppl}. The correlation function $\langle \hat{n}_{j,A}(t) \hat{n}_{0,A}\rangle $ is measured at every time slice as we progressively evolve $n_{0,A}|\psi_0\rangle$ with operator $e^{-iH t_l}$ in TDVP. For $N_t$, $n_{0,A}$ and the details of spectrum calculation, see SM~\cite{suppl}.
We also measure the single-particle Green's functions $\langle b_{j,A}(t_l) b^\dagger_{0,A}\rangle$ and $\langle b^\dagger_{j,A}(t_l) b_{0,A}\rangle$ respectively and label the corresponding spectra with $G_e^A(\mathbf{k},\omega)$ (electron) and $G_h^A(\mathbf{k},\omega)$ (hole).
We sum $G_e^A(\mathbf{k},\omega)$ and $G_h^A(\mathbf{-k},-\omega)$ to obtain single-particle spectrum $G^A(\mathbf{k},\omega)$, where we use the particle-hole conjugation relation to reverse the sign of momentum and energy for holes. $b^\dagger$ and $b$ locate at the bulk of the cylinder. The local density of state can then be obtained as $\rho(\omega) = \frac{1}{N_k} \sum_{\mathbf{k}}G^A(\mathbf{k},\omega) $. 

\begin{figure}[htp!]
    \centering
    \includegraphics[width=0.7\columnwidth]{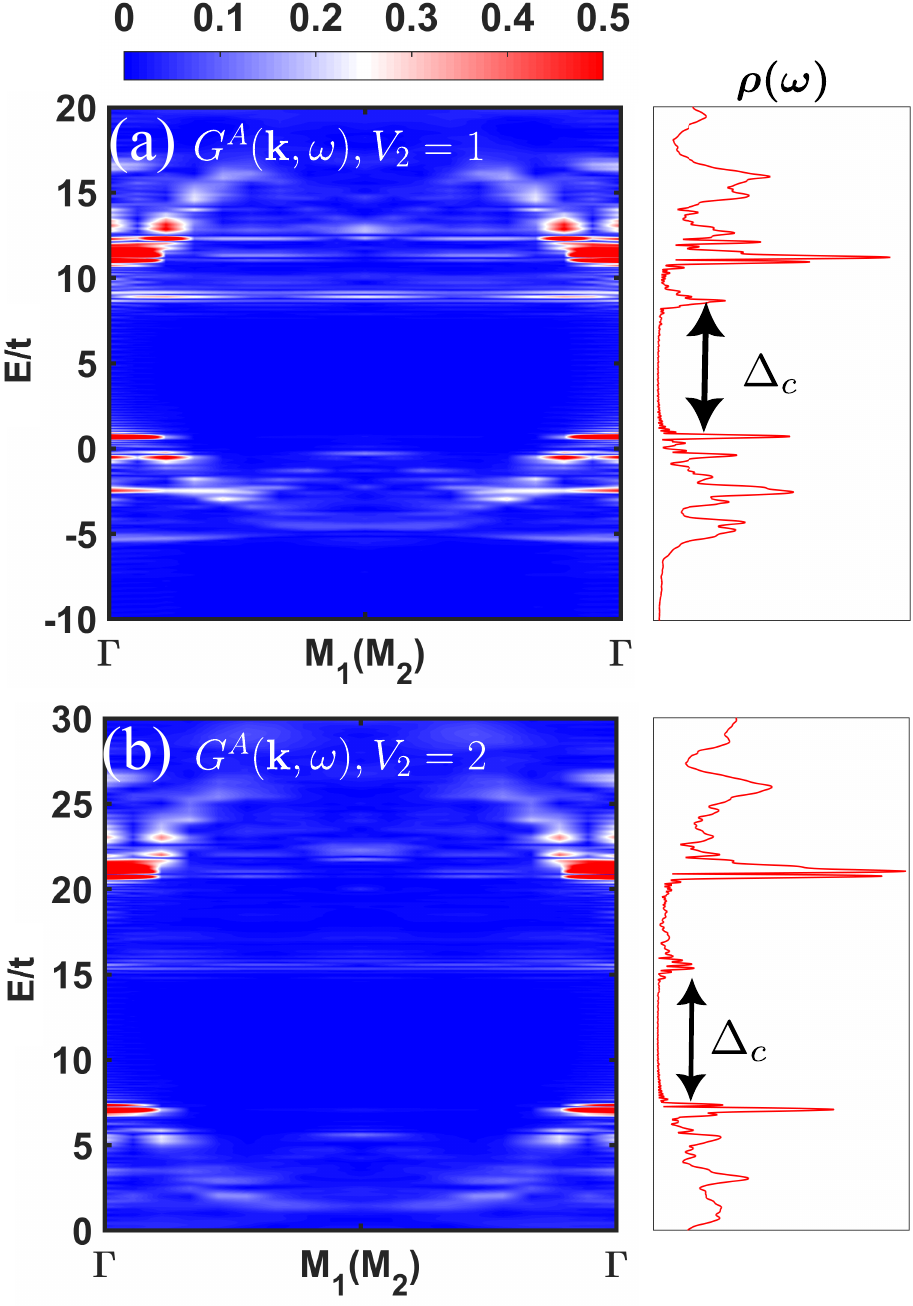}
\caption{\textbf{Single-particle spectrum inside the FCI phase.} (a), (b) are the $G^A(\mathbf{k},\omega)$ and $\rho(\omega)$ inside the FCI phase (with the same parameter as Fig.~\ref{fig:fig2} (a) and (b)) and close to the FCI-Solid I transition (with the same parameter as Fig.~\ref{fig:fig2} (c) and (d)), respectively. In both cases, the single-particle gap $\Delta_c \approx 7$ are much larger than the roton gaps $\Delta_{n_1}$ and $\Delta_{n_2}$. The path (Path I) we have chosen to display the single-particle spectrum features a minimal charge gap and encompasses major of the total spectral weight.}
    \label{fig:fig3}
\end{figure}

Fig.~\ref{fig:fig2} shows the $S^{A}(\mathbf{k},\omega)$ along the path I and III in the BZ in Fig.~\ref{fig:fig1} (d). Panels (a) and (b) of Fig.~\ref{fig:fig2} are deep inside the FCI phase, and one clearly sees the lowest roton minimal at $M_1$ with the roton gap $\Delta_{n_1} \sim 1$, which is much smaller the single-particle gap computed at the same parameter  which is $\Delta_c \sim 7$ with sharper spectrum weight, as shown in Fig.~\ref{fig:fig3} (a). This observation further reinforces the understanding that in the FCI phase, the lowest excitations are the charge neutral roton instead of the single-particle ones~\cite{Lu2024Thermodynamics,LuFractional2024,luFrom2024}. Panel (b) exhibits the similar lowest rotons at $M_3$ and $M_5$, but more interestingly, there is another higher energy roton manifesting at momentum $\mathbf{b_1}/4+\mathbf{b_2}/2$, this is the second roton that will trigger the vestigial transition from Solid I to Solid II, once the lowest roton $\Delta_{n_1}$ condense to give rise to the Solid I. In Fig.~\ref{fig:fig2} (c) and (d), the parameter $(V_1=4, V_2=2)$ is closer to the FCI-Solid I transition, and one sees that the roton gap $\Delta_{n_1}$ becomes smaller, signifying the approaching transition. For the discussion of magnetoroton in $1/3$ fermionic FCI, please refer to SM~\cite{suppl}.

\updated{We note that as the roton is about to condense, its spectral weight become much sharper and concentrated close to its minimum compared with that deep inside the FCI state, we expect such clear dynamic response of the magnetoroton of FCI close to its transition to CDW phase, can be seen in the future Raman scattering experiments. 
Moreover, our previous finite-temperature calculations have shown the thermodynamic signals of the neutral excitations of (F)QAH states in the correlated flat band systems. As the temperature approaches the energy scale of the neutral gap, the compressibility of the system is greatly enhanced~\cite{panThermodynamic2023,Lu2024Thermodynamics,luFrom2024,luVestigial2024,excitonLin2022,luGeneric2025}},.

We also monitor the evolution of the single-particle spectra along the same parameter path, and the results are shown in Fig.~\ref{fig:fig3}. We plot $G^A(\mathbf{k},\omega)$ and $\rho(\omega)$ in Fig.~\ref{fig:fig3} (a) with $V_2=1$ (well inside the FCI phase) and Fig.~\ref{fig:fig3} (b) with $V_2=2$ (close to the FCI-Solid I transition). It is clear that at both parameters, the single-particle gaps are well developed and much larger than the roton gaps at the same parameter. Such results suggest that, the transition out of a FCI can be purely driven by the charge-neutral roton modes, not related the single-particle gap~\cite{Lu2024Interaction,luFrom2024,LuFractional2024,luVestigial2024}.

As shown in Fig.~\ref{fig:fig2} (a) and (c), the chiral graviton mode (CGM) is invisible in $ S^A(\mathbf{k},\omega)$, to investigate CGM in our FCI system, we proceed to compute the dynamic quadrupolar correlation function, as we now turn to.

\noindent{\textcolor{blue}{\it Chiral Gravitons.}---} \updated{We combine the quadrupolar density operator with chiral form factor to probe CGM  phenomenologically. We define $O^{\pm}_m=\sum_l e^{\pm iL_z\phi_l}n_{r_m}  n_{{r_{m+\delta_l}}}$, where $l \in \{1,2,\dots,6\}$ connect site $m$ with it's 6 next nearest neighbor on same sub-lattice and $\phi_l = l\frac{\pi}{3}$ with the $l$ runs counterclockwise~\cite{Wang2022,Liou2019Chiral}, such that $O^{\pm}_m$ create a chiral excitation with angular momentum $\pm 2$ at site $m$.  We therefore define the correlation function of $O^{\pm}_m$}
\begin{equation}
    \begin{aligned}
        G^{\pm}(\mathbf{k},\omega) & =\frac{1}{\sqrt{N_tN_A}} \sum_{jl} e^{i(\omega+i\eta) t_l} e^{-i\mathbf{k}\cdot(\mathbf{x_j}-\mathbf{x_0})}  \langle O^{\mp}_{j}(t_l)  O^{\pm}_{0}\rangle 
        \label{eq:gravitoncoor}
    \end{aligned}
\end{equation}
\updated{with $\eta = 0.05$, $(O^{\pm})^\dagger =O^{\mp}$, In the calculation of Eq.~\eqref{eq:gravitoncoor}, the summation over the sites runs over the same sublattice. The chirality of the CGM is determined by comparing the spectrum weight of $G^{\pm}(\mathbf{k},\omega)$ at $\Gamma$ point with $G^+(\mathbf{k},\omega)$ probes $S = -2$ signal vise versa). We further combine $g^+(\mathbf{k},\omega) = G^{+}(\mathbf{k},\omega) - G^{-}(\mathbf{k},\omega)$ and $g^-(\mathbf{k},\omega) = G^{-}(\mathbf{k},\omega) - G^{+}(\mathbf{k},\omega)$ such that their subtraction cancels the non-chiral part of the spectral signature. We note that $g^\pm(\mathbf{k},\omega)$ is also the time-ordered Green's function of $O^{\pm}_m$.}

\begin{figure}[ht!]
    \centering
    \includegraphics[width=\columnwidth]{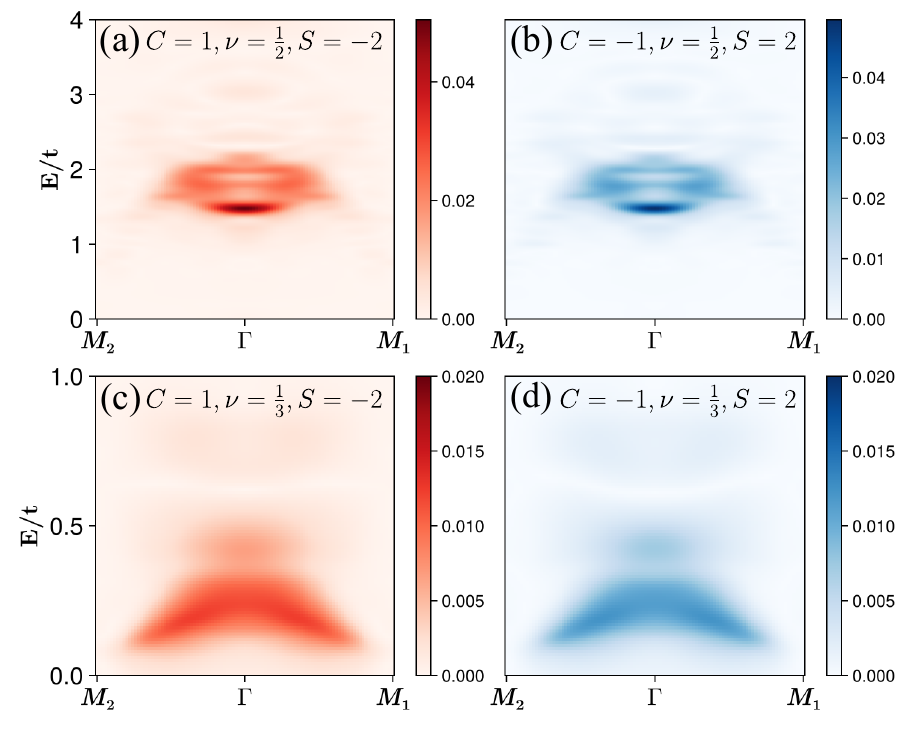}
    \caption{\textbf{Chiral graviton mode (CGM) inside FCI phase and its tunability}.  (a) and is the dynamic quadrupolar density correlations $g^-$ ($S = -2$)  measured at $V_1 = 4, V_2 = 1$. The chiral $S = -2$ graviton extends around $\Gamma$ point. In (b), we revert the magnetic flux $\phi$ such that the Chern number for the lower band is $C=-1$ and calculate $g^+$ as in (a). (c) and (d) are the same measurement as (a) and (b), but on $1/3$ FCI with $V_1 = 1.5, V_2 = 0$, where the chirality reversion is also observed}
    \label{fig:fig4}
\end{figure}

\updated{Fig.~\ref{fig:fig4} (a) is the $g^{-}(\mathbf{k},\omega)$ at $V_1=4, V_2=1$ inside the $1/2$ bosonic FCI phase. One sees a strong charge-neutral response close to the momentum $\Gamma$ at $(S=-2)$ sector with energy twice that of the magnetoroton. This is also the momentum where the CGM was proposed in the FQHE literature~\cite{Yang_2012,Golkar2016,Yuzhu2023Geometric,Yuzhu2025Dynamics,Liu2024}, and experiment~\cite{Liang_2024}.  One more interesting analysis we did here is to switch the Chern number and chirality of the flat band in our model, this is similar with the observation of the CGM in the FQHE experiment that hole doping FQHE state ($\nu=2/3$) have opposite chirality compared with it's electron ($\nu=1/3$) counterpart or flip the direction of the magnetic field. As shown in Fig.~\ref{fig:fig4} (b), once we revert the flux directions in the in the $t'$-term in Eq.~\eqref{eq:eq1}, at the same interaction strength of FCI phase, the graviton appears near the $\Gamma$ point in $g^{+}(\mathbf{k},\omega)$. The result for $1/3$ fermionic FCI with $V_1 = 1.5, V_2 = 0$ is shown in Fig.~\ref{fig:fig4} (c)(d), the spectrum peak is located at the $\Gamma$ point, and the chiral mode switching is observed upon reversing the flux directions. However, the chiral signal is weaker and of distinct shape compared with the bosonic case. For bosonic FCI, we think such tunability of the chirality of the graviton in the FCI system could be potentially observed in the circularly polarized resonant inelastic light scattering experiment ~\cite{hirjibehedinSplitting2005,Liang_2024}. For fermionic FCI, we also expect the tunability of the chirality for CGM.}

\updated{We also trace the evolution of the spectrum weight upon increasing cylinder length. The result is shown in Section II of the SM\cite{suppl}. The spectrum peak gradually declines at larger system sizes. This trend has been observed on both FQH\cite{Liou2019} and FCI states\cite{Yuzhu2025Dynamics}, which is attributed to the increase of density of states at larger systems, introducing more scattering channels and consequently lowering the lifetime of CGM\cite{Yuzhu2025Dynamics}. Although our phenomenological operator has a much wider momentum spread compared with the bosonic case. We leave the clearer identification of the CGM in fermionic FCI models and its lifetime in the thermodynamic limit for future investigations.}

\noindent{\textcolor{blue}{\it Discussion and conclusions.}---} In this work, we observe the clear spectral properties of the magnetoroton and chiral graviton, as the collective charge-neutral excitations in FCI systems. We discover that the roton minimum can go soft and give rise to a CDW phase (with the corresponding wavevector), while during the process, the spectral weight of the roton excitation becomes sharper and more concentrated, while the single-particle gap remains large. The enhanced spectral signature is expected to be measured in Raman scattering. 
Moreover, we not only show that the graviton mode is chiral for a given FCI state, but also directly visualize the different chiralities of the CGM of FCIs by tuning the sign of the Hall conductance
and such novel phenomena can potentially be seen via the resonant inelastic light scattering, as has recently been seen in the FQH experiment~\cite{Liang_2024}. 

{\it Acknowledgments}\,---\, We thank Yang Liu, Yuzhu Wang, Bin-Bin Chen, Bo Yang, Wei Zhu, Wang Yao and Di Xiao for helpful discussions. 
ML, HYL and ZYM acknowledge the support from the Research Grants Council (RGC) of Hong Kong (Project Nos. AoE/P-701/20, 17309822, HKU C7037-22GF, 17302223, 17301924), the ANR/RGC Joint Research Scheme sponsored by RGC of Hong Kong and French National Research Agency (Project No. A\_HKU703/22) and the HKU Seed Funding for Strategic Interdisciplinary Research “Many-body paradigm in
quantum moiré material research”. We thank HPC2021 system under the Information Technology Services and the Blackbody HPC system at the Department of Physics, University of Hong Kong, as well as the Beijng PARATERA Tech CO.,Ltd. (URL: https://cloud.paratera.com) for providing HPC resources that have contributed to the research results reported within this paper. H.Q. Wu acknowledge the
support from the National Natural Science Foundation of China (Grant No. 12474248), GuangDong Basic and Applied Basic Research Foundation (Grant No. 2023B1515120013).

{\it Note Added}\,---\, Upon the completion of our work, we became aware of Ref.~\cite{shenMagnetorotons2024} which described the neutral excitations of FCI in the continuum model by exact diagonalizations and SMA.

\bibliographystyle{apsrev4-2}
\bibliography{temp_bibtex}

\begin{thebibliography}{68}%
\makeatletter
\providecommand \@ifxundefined [1]{%
 \@ifx{#1\undefined}
}%
\providecommand \@ifnum [1]{%
 \ifnum #1\expandafter \@firstoftwo
 \else \expandafter \@secondoftwo
 \fi
}%
\providecommand \@ifx [1]{%
 \ifx #1\expandafter \@firstoftwo
 \else \expandafter \@secondoftwo
 \fi
}%
\providecommand \natexlab [1]{#1}%
\providecommand \enquote  [1]{``#1''}%
\providecommand \bibnamefont  [1]{#1}%
\providecommand \bibfnamefont [1]{#1}%
\providecommand \citenamefont [1]{#1}%
\providecommand \href@noop [0]{\@secondoftwo}%
\providecommand \href [0]{\begingroup \@sanitize@url \@href}%
\providecommand \@href[1]{\@@startlink{#1}\@@href}%
\providecommand \@@href[1]{\endgroup#1\@@endlink}%
\providecommand \@sanitize@url [0]{\catcode `\\12\catcode `\$12\catcode `\&12\catcode `\#12\catcode `\^12\catcode `\_12\catcode `\%12\relax}%
\providecommand \@@startlink[1]{}%
\providecommand \@@endlink[0]{}%
\providecommand \url  [0]{\begingroup\@sanitize@url \@url }%
\providecommand \@url [1]{\endgroup\@href {#1}{\urlprefix }}%
\providecommand \urlprefix  [0]{URL }%
\providecommand \Eprint [0]{\href }%
\providecommand \doibase [0]{https://doi.org/}%
\providecommand \selectlanguage [0]{\@gobble}%
\providecommand \bibinfo  [0]{\@secondoftwo}%
\providecommand \bibfield  [0]{\@secondoftwo}%
\providecommand \translation [1]{[#1]}%
\providecommand \BibitemOpen [0]{}%
\providecommand \bibitemStop [0]{}%
\providecommand \bibitemNoStop [0]{.\EOS\space}%
\providecommand \EOS [0]{\spacefactor3000\relax}%
\providecommand \BibitemShut  [1]{\csname bibitem#1\endcsname}%
\let\auto@bib@innerbib\@empty
\bibitem [{\citenamefont {Girvin}\ \emph {et~al.}(1985)\citenamefont {Girvin}, \citenamefont {MacDonald},\ and\ \citenamefont {Platzman}}]{Girvin_1985}%
  \BibitemOpen
  \bibfield  {author} {\bibinfo {author} {\bibfnamefont {S.~M.}\ \bibnamefont {Girvin}}, \bibinfo {author} {\bibfnamefont {A.~H.}\ \bibnamefont {MacDonald}},\ and\ \bibinfo {author} {\bibfnamefont {P.~M.}\ \bibnamefont {Platzman}},\ }\href {https://doi.org/10.1103/physrevlett.54.581} {\bibfield  {journal} {\bibinfo  {journal} {Physical Review Letters}\ }\textbf {\bibinfo {volume} {54}},\ \bibinfo {pages} {581–583} (\bibinfo {year} {1985})}\BibitemShut {NoStop}%
\bibitem [{\citenamefont {Scarola}\ \emph {et~al.}(2000)\citenamefont {Scarola}, \citenamefont {Park},\ and\ \citenamefont {Jain}}]{scarola2000rotons}%
  \BibitemOpen
  \bibfield  {author} {\bibinfo {author} {\bibfnamefont {V.~W.}\ \bibnamefont {Scarola}}, \bibinfo {author} {\bibfnamefont {K.}~\bibnamefont {Park}},\ and\ \bibinfo {author} {\bibfnamefont {J.~K.}\ \bibnamefont {Jain}},\ }\href@noop {} {\bibfield  {journal} {\bibinfo  {journal} {Physical Review B}\ }\textbf {\bibinfo {volume} {61}},\ \bibinfo {pages} {13064} (\bibinfo {year} {2000})}\BibitemShut {NoStop}%
\bibitem [{\citenamefont {Du}\ \emph {et~al.}(1993)\citenamefont {Du}, \citenamefont {Stormer}, \citenamefont {Tsui}, \citenamefont {Pfeiffer},\ and\ \citenamefont {West}}]{Du1993Experimental}%
  \BibitemOpen
  \bibfield  {author} {\bibinfo {author} {\bibfnamefont {R.~R.}\ \bibnamefont {Du}}, \bibinfo {author} {\bibfnamefont {H.~L.}\ \bibnamefont {Stormer}}, \bibinfo {author} {\bibfnamefont {D.~C.}\ \bibnamefont {Tsui}}, \bibinfo {author} {\bibfnamefont {L.~N.}\ \bibnamefont {Pfeiffer}},\ and\ \bibinfo {author} {\bibfnamefont {K.~W.}\ \bibnamefont {West}},\ }\href {https://doi.org/10.1103/PhysRevLett.70.2944} {\bibfield  {journal} {\bibinfo  {journal} {Phys. Rev. Lett.}\ }\textbf {\bibinfo {volume} {70}},\ \bibinfo {pages} {2944} (\bibinfo {year} {1993})}\BibitemShut {NoStop}%
\bibitem [{\citenamefont {Pinczuk}\ \emph {et~al.}(1993)\citenamefont {Pinczuk}, \citenamefont {Dennis}, \citenamefont {Pfeiffer},\ and\ \citenamefont {West}}]{Pinczuk1993Observation}%
  \BibitemOpen
  \bibfield  {author} {\bibinfo {author} {\bibfnamefont {A.}~\bibnamefont {Pinczuk}}, \bibinfo {author} {\bibfnamefont {B.~S.}\ \bibnamefont {Dennis}}, \bibinfo {author} {\bibfnamefont {L.~N.}\ \bibnamefont {Pfeiffer}},\ and\ \bibinfo {author} {\bibfnamefont {K.}~\bibnamefont {West}},\ }\href {https://doi.org/10.1103/PhysRevLett.70.3983} {\bibfield  {journal} {\bibinfo  {journal} {Phys. Rev. Lett.}\ }\textbf {\bibinfo {volume} {70}},\ \bibinfo {pages} {3983} (\bibinfo {year} {1993})}\BibitemShut {NoStop}%
\bibitem [{\citenamefont {Kang}\ \emph {et~al.}(2001)\citenamefont {Kang}, \citenamefont {Pinczuk}, \citenamefont {Dennis}, \citenamefont {Pfeiffer},\ and\ \citenamefont {West}}]{Kang2001Observation}%
  \BibitemOpen
  \bibfield  {author} {\bibinfo {author} {\bibfnamefont {M.}~\bibnamefont {Kang}}, \bibinfo {author} {\bibfnamefont {A.}~\bibnamefont {Pinczuk}}, \bibinfo {author} {\bibfnamefont {B.~S.}\ \bibnamefont {Dennis}}, \bibinfo {author} {\bibfnamefont {L.~N.}\ \bibnamefont {Pfeiffer}},\ and\ \bibinfo {author} {\bibfnamefont {K.~W.}\ \bibnamefont {West}},\ }\href {https://doi.org/10.1103/PhysRevLett.86.2637} {\bibfield  {journal} {\bibinfo  {journal} {Phys. Rev. Lett.}\ }\textbf {\bibinfo {volume} {86}},\ \bibinfo {pages} {2637} (\bibinfo {year} {2001})}\BibitemShut {NoStop}%
\bibitem [{\citenamefont {Kukushkin}\ \emph {et~al.}(2009)\citenamefont {Kukushkin}, \citenamefont {Smet}, \citenamefont {Scarola}, \citenamefont {Umansky},\ and\ \citenamefont {von Klitzing}}]{Igor2009Dispersion}%
  \BibitemOpen
  \bibfield  {author} {\bibinfo {author} {\bibfnamefont {I.~V.}\ \bibnamefont {Kukushkin}}, \bibinfo {author} {\bibfnamefont {J.~H.}\ \bibnamefont {Smet}}, \bibinfo {author} {\bibfnamefont {V.~W.}\ \bibnamefont {Scarola}}, \bibinfo {author} {\bibfnamefont {V.}~\bibnamefont {Umansky}},\ and\ \bibinfo {author} {\bibfnamefont {K.}~\bibnamefont {von Klitzing}},\ }\href {https://doi.org/10.1126/science.1171472} {\bibfield  {journal} {\bibinfo  {journal} {Science}\ }\textbf {\bibinfo {volume} {324}},\ \bibinfo {pages} {1044} (\bibinfo {year} {2009})},\ \Eprint {https://arxiv.org/abs/https://www.science.org/doi/pdf/10.1126/science.1171472} {https://www.science.org/doi/pdf/10.1126/science.1171472} \BibitemShut {NoStop}%
\bibitem [{\citenamefont {Kumar}\ and\ \citenamefont {Bhatt}(2022)}]{Kumar_2022}%
  \BibitemOpen
  \bibfield  {author} {\bibinfo {author} {\bibfnamefont {P.}~\bibnamefont {Kumar}}\ and\ \bibinfo {author} {\bibfnamefont {R.~N.}\ \bibnamefont {Bhatt}},\ }\href {https://doi.org/10.1103/PhysRevB.106.115101} {\bibfield  {journal} {\bibinfo  {journal} {Phys. Rev. B}\ }\textbf {\bibinfo {volume} {106}},\ \bibinfo {pages} {115101} (\bibinfo {year} {2022})}\BibitemShut {NoStop}%
\bibitem [{\citenamefont {Tsui}\ \emph {et~al.}(2024)\citenamefont {Tsui}, \citenamefont {He}, \citenamefont {Hu}, \citenamefont {Lake}, \citenamefont {Wang}, \citenamefont {Watanabe}, \citenamefont {Taniguchi}, \citenamefont {Zaletel},\ and\ \citenamefont {Yazdani}}]{tsui2024direct}%
  \BibitemOpen
  \bibfield  {author} {\bibinfo {author} {\bibfnamefont {Y.-C.}\ \bibnamefont {Tsui}}, \bibinfo {author} {\bibfnamefont {M.}~\bibnamefont {He}}, \bibinfo {author} {\bibfnamefont {Y.}~\bibnamefont {Hu}}, \bibinfo {author} {\bibfnamefont {E.}~\bibnamefont {Lake}}, \bibinfo {author} {\bibfnamefont {T.}~\bibnamefont {Wang}}, \bibinfo {author} {\bibfnamefont {K.}~\bibnamefont {Watanabe}}, \bibinfo {author} {\bibfnamefont {T.}~\bibnamefont {Taniguchi}}, \bibinfo {author} {\bibfnamefont {M.~P.}\ \bibnamefont {Zaletel}},\ and\ \bibinfo {author} {\bibfnamefont {A.}~\bibnamefont {Yazdani}},\ }\href@noop {} {\bibfield  {journal} {\bibinfo  {journal} {Nature}\ }\textbf {\bibinfo {volume} {628}},\ \bibinfo {pages} {287} (\bibinfo {year} {2024})}\BibitemShut {NoStop}%
\bibitem [{\citenamefont {Song}\ \emph {et~al.}(2024)\citenamefont {Song}, \citenamefont {Jian}, \citenamefont {Fu},\ and\ \citenamefont {Xu}}]{song2024intertwined}%
  \BibitemOpen
  \bibfield  {author} {\bibinfo {author} {\bibfnamefont {X.-Y.}\ \bibnamefont {Song}}, \bibinfo {author} {\bibfnamefont {C.-M.}\ \bibnamefont {Jian}}, \bibinfo {author} {\bibfnamefont {L.}~\bibnamefont {Fu}},\ and\ \bibinfo {author} {\bibfnamefont {C.}~\bibnamefont {Xu}},\ }\href@noop {} {\bibfield  {journal} {\bibinfo  {journal} {Physical Review B}\ }\textbf {\bibinfo {volume} {109}},\ \bibinfo {pages} {115116} (\bibinfo {year} {2024})}\BibitemShut {NoStop}%
\bibitem [{\citenamefont {Haldane}(2011)}]{Haldane2011}%
  \BibitemOpen
  \bibfield  {author} {\bibinfo {author} {\bibfnamefont {F.~D.~M.}\ \bibnamefont {Haldane}},\ }\href {https://doi.org/10.1103/PhysRevLett.107.116801} {\bibfield  {journal} {\bibinfo  {journal} {Phys. Rev. Lett.}\ }\textbf {\bibinfo {volume} {107}},\ \bibinfo {pages} {116801} (\bibinfo {year} {2011})}\BibitemShut {NoStop}%
\bibitem [{\citenamefont {Qiu}\ \emph {et~al.}(2012)\citenamefont {Qiu}, \citenamefont {Haldane}, \citenamefont {Wan}, \citenamefont {Yang},\ and\ \citenamefont {Yi}}]{Qiu2012}%
  \BibitemOpen
  \bibfield  {author} {\bibinfo {author} {\bibfnamefont {R.-Z.}\ \bibnamefont {Qiu}}, \bibinfo {author} {\bibfnamefont {F.~D.~M.}\ \bibnamefont {Haldane}}, \bibinfo {author} {\bibfnamefont {X.}~\bibnamefont {Wan}}, \bibinfo {author} {\bibfnamefont {K.}~\bibnamefont {Yang}},\ and\ \bibinfo {author} {\bibfnamefont {S.}~\bibnamefont {Yi}},\ }\href {https://doi.org/10.1103/PhysRevB.85.115308} {\bibfield  {journal} {\bibinfo  {journal} {Phys. Rev. B}\ }\textbf {\bibinfo {volume} {85}},\ \bibinfo {pages} {115308} (\bibinfo {year} {2012})}\BibitemShut {NoStop}%
\bibitem [{\citenamefont {Liou}\ \emph {et~al.}(2019{\natexlab{a}})\citenamefont {Liou}, \citenamefont {Haldane}, \citenamefont {Yang},\ and\ \citenamefont {Rezayi}}]{Liou2019}%
  \BibitemOpen
  \bibfield  {author} {\bibinfo {author} {\bibfnamefont {S.-F.}\ \bibnamefont {Liou}}, \bibinfo {author} {\bibfnamefont {F.~D.~M.}\ \bibnamefont {Haldane}}, \bibinfo {author} {\bibfnamefont {K.}~\bibnamefont {Yang}},\ and\ \bibinfo {author} {\bibfnamefont {E.~H.}\ \bibnamefont {Rezayi}},\ }\href {https://doi.org/10.1103/PhysRevLett.123.146801} {\bibfield  {journal} {\bibinfo  {journal} {Phys. Rev. Lett.}\ }\textbf {\bibinfo {volume} {123}},\ \bibinfo {pages} {146801} (\bibinfo {year} {2019}{\natexlab{a}})}\BibitemShut {NoStop}%
\bibitem [{\citenamefont {Johri}\ \emph {et~al.}(2016)\citenamefont {Johri}, \citenamefont {Papić}, \citenamefont {Schmitteckert}, \citenamefont {Bhatt},\ and\ \citenamefont {Haldane}}]{Johri2016}%
  \BibitemOpen
  \bibfield  {author} {\bibinfo {author} {\bibfnamefont {S.}~\bibnamefont {Johri}}, \bibinfo {author} {\bibfnamefont {Z.}~\bibnamefont {Papić}}, \bibinfo {author} {\bibfnamefont {P.}~\bibnamefont {Schmitteckert}}, \bibinfo {author} {\bibfnamefont {R.~N.}\ \bibnamefont {Bhatt}},\ and\ \bibinfo {author} {\bibfnamefont {F.~D.~M.}\ \bibnamefont {Haldane}},\ }\href {https://doi.org/10.1088/1367-2630/18/2/025011} {\bibfield  {journal} {\bibinfo  {journal} {New Journal of Physics}\ }\textbf {\bibinfo {volume} {18}},\ \bibinfo {pages} {025011} (\bibinfo {year} {2016})}\BibitemShut {NoStop}%
\bibitem [{\citenamefont {Haldane}\ \emph {et~al.}(2021)\citenamefont {Haldane}, \citenamefont {Rezayi},\ and\ \citenamefont {Yang}}]{Haldane2021}%
  \BibitemOpen
  \bibfield  {author} {\bibinfo {author} {\bibfnamefont {F.~D.~M.}\ \bibnamefont {Haldane}}, \bibinfo {author} {\bibfnamefont {E.~H.}\ \bibnamefont {Rezayi}},\ and\ \bibinfo {author} {\bibfnamefont {K.}~\bibnamefont {Yang}},\ }\href {https://doi.org/10.1103/PhysRevB.104.L121106} {\bibfield  {journal} {\bibinfo  {journal} {Phys. Rev. B}\ }\textbf {\bibinfo {volume} {104}},\ \bibinfo {pages} {L121106} (\bibinfo {year} {2021})}\BibitemShut {NoStop}%
\bibitem [{\citenamefont {Liou}\ \emph {et~al.}(2019{\natexlab{b}})\citenamefont {Liou}, \citenamefont {Haldane}, \citenamefont {Yang},\ and\ \citenamefont {Rezayi}}]{Liou2019Chiral}%
  \BibitemOpen
  \bibfield  {author} {\bibinfo {author} {\bibfnamefont {S.-F.}\ \bibnamefont {Liou}}, \bibinfo {author} {\bibfnamefont {F.~D.~M.}\ \bibnamefont {Haldane}}, \bibinfo {author} {\bibfnamefont {K.}~\bibnamefont {Yang}},\ and\ \bibinfo {author} {\bibfnamefont {E.~H.}\ \bibnamefont {Rezayi}},\ }\href {https://doi.org/10.1103/PhysRevLett.123.146801} {\bibfield  {journal} {\bibinfo  {journal} {Phys. Rev. Lett.}\ }\textbf {\bibinfo {volume} {123}},\ \bibinfo {pages} {146801} (\bibinfo {year} {2019}{\natexlab{b}})}\BibitemShut {NoStop}%
\bibitem [{\citenamefont {Wang}\ and\ \citenamefont {Yang}(2022)}]{Wang2022}%
  \BibitemOpen
  \bibfield  {author} {\bibinfo {author} {\bibfnamefont {Y.}~\bibnamefont {Wang}}\ and\ \bibinfo {author} {\bibfnamefont {B.}~\bibnamefont {Yang}},\ }\href {https://doi.org/10.1103/PhysRevB.105.035144} {\bibfield  {journal} {\bibinfo  {journal} {Phys. Rev. B}\ }\textbf {\bibinfo {volume} {105}},\ \bibinfo {pages} {035144} (\bibinfo {year} {2022})}\BibitemShut {NoStop}%
\bibitem [{\citenamefont {Yang}\ \emph {et~al.}(2012)\citenamefont {Yang}, \citenamefont {Hu}, \citenamefont {Papi\ifmmode~\acute{c}\else \'{c}\fi{}},\ and\ \citenamefont {Haldane}}]{Yang_2012}%
  \BibitemOpen
  \bibfield  {author} {\bibinfo {author} {\bibfnamefont {B.}~\bibnamefont {Yang}}, \bibinfo {author} {\bibfnamefont {Z.-X.}\ \bibnamefont {Hu}}, \bibinfo {author} {\bibfnamefont {Z.}~\bibnamefont {Papi\ifmmode~\acute{c}\else \'{c}\fi{}}},\ and\ \bibinfo {author} {\bibfnamefont {F.~D.~M.}\ \bibnamefont {Haldane}},\ }\href {https://doi.org/10.1103/PhysRevLett.108.256807} {\bibfield  {journal} {\bibinfo  {journal} {Phys. Rev. Lett.}\ }\textbf {\bibinfo {volume} {108}},\ \bibinfo {pages} {256807} (\bibinfo {year} {2012})}\BibitemShut {NoStop}%
\bibitem [{\citenamefont {Liang}\ \emph {et~al.}(2024)\citenamefont {Liang}, \citenamefont {Liu}, \citenamefont {Yang}, \citenamefont {Huang}, \citenamefont {Wurstbauer}, \citenamefont {Dean}, \citenamefont {West}, \citenamefont {Pfeiffer}, \citenamefont {Du},\ and\ \citenamefont {Pinczuk}}]{Liang_2024}%
  \BibitemOpen
  \bibfield  {author} {\bibinfo {author} {\bibfnamefont {J.}~\bibnamefont {Liang}}, \bibinfo {author} {\bibfnamefont {Z.}~\bibnamefont {Liu}}, \bibinfo {author} {\bibfnamefont {Z.}~\bibnamefont {Yang}}, \bibinfo {author} {\bibfnamefont {Y.}~\bibnamefont {Huang}}, \bibinfo {author} {\bibfnamefont {U.}~\bibnamefont {Wurstbauer}}, \bibinfo {author} {\bibfnamefont {C.~R.}\ \bibnamefont {Dean}}, \bibinfo {author} {\bibfnamefont {K.~W.}\ \bibnamefont {West}}, \bibinfo {author} {\bibfnamefont {L.~N.}\ \bibnamefont {Pfeiffer}}, \bibinfo {author} {\bibfnamefont {L.}~\bibnamefont {Du}},\ and\ \bibinfo {author} {\bibfnamefont {A.}~\bibnamefont {Pinczuk}},\ }\href {https://doi.org/10.1038/s41586-024-07201-w} {\bibfield  {journal} {\bibinfo  {journal} {Nature}\ }\textbf {\bibinfo {volume} {628}},\ \bibinfo {pages} {78–83} (\bibinfo {year} {2024})}\BibitemShut {NoStop}%
\bibitem [{\citenamefont {Xia}\ \emph {et~al.}(2011)\citenamefont {Xia}, \citenamefont {Eisenstein}, \citenamefont {Pfeiffer},\ and\ \citenamefont {West}}]{Xia_2011}%
  \BibitemOpen
  \bibfield  {author} {\bibinfo {author} {\bibfnamefont {J.}~\bibnamefont {Xia}}, \bibinfo {author} {\bibfnamefont {J.~P.}\ \bibnamefont {Eisenstein}}, \bibinfo {author} {\bibfnamefont {L.~N.}\ \bibnamefont {Pfeiffer}},\ and\ \bibinfo {author} {\bibfnamefont {K.~W.}\ \bibnamefont {West}},\ }\href {https://doi.org/10.1038/nphys2118} {\bibfield  {journal} {\bibinfo  {journal} {Nature Physics}\ }\textbf {\bibinfo {volume} {7}},\ \bibinfo {pages} {845–848} (\bibinfo {year} {2011})}\BibitemShut {NoStop}%
\bibitem [{\citenamefont {Yang}(2020)}]{Yang_2020}%
  \BibitemOpen
  \bibfield  {author} {\bibinfo {author} {\bibfnamefont {B.}~\bibnamefont {Yang}},\ }\href {https://doi.org/10.1103/PhysRevResearch.2.033362} {\bibfield  {journal} {\bibinfo  {journal} {Phys. Rev. Res.}\ }\textbf {\bibinfo {volume} {2}},\ \bibinfo {pages} {033362} (\bibinfo {year} {2020})}\BibitemShut {NoStop}%
\bibitem [{\citenamefont {Pu}\ \emph {et~al.}(2024)\citenamefont {Pu}, \citenamefont {Balram}, \citenamefont {Taylor}, \citenamefont {Fradkin},\ and\ \citenamefont {Papi\ifmmode~\acute{c}\else \'{c}\fi{}}}]{Pu_2024}%
  \BibitemOpen
  \bibfield  {author} {\bibinfo {author} {\bibfnamefont {S.}~\bibnamefont {Pu}}, \bibinfo {author} {\bibfnamefont {A.~C.}\ \bibnamefont {Balram}}, \bibinfo {author} {\bibfnamefont {J.}~\bibnamefont {Taylor}}, \bibinfo {author} {\bibfnamefont {E.}~\bibnamefont {Fradkin}},\ and\ \bibinfo {author} {\bibfnamefont {Z.}~\bibnamefont {Papi\ifmmode~\acute{c}\else \'{c}\fi{}}},\ }\href {https://doi.org/10.1103/PhysRevLett.132.236503} {\bibfield  {journal} {\bibinfo  {journal} {Phys. Rev. Lett.}\ }\textbf {\bibinfo {volume} {132}},\ \bibinfo {pages} {236503} (\bibinfo {year} {2024})}\BibitemShut {NoStop}%
\bibitem [{\citenamefont {Regnault}\ \emph {et~al.}(2017)\citenamefont {Regnault}, \citenamefont {Maciejko}, \citenamefont {Kivelson},\ and\ \citenamefont {Sondhi}}]{Regnault_2017}%
  \BibitemOpen
  \bibfield  {author} {\bibinfo {author} {\bibfnamefont {N.}~\bibnamefont {Regnault}}, \bibinfo {author} {\bibfnamefont {J.}~\bibnamefont {Maciejko}}, \bibinfo {author} {\bibfnamefont {S.~A.}\ \bibnamefont {Kivelson}},\ and\ \bibinfo {author} {\bibfnamefont {S.~L.}\ \bibnamefont {Sondhi}},\ }\href {https://doi.org/10.1103/PhysRevB.96.035150} {\bibfield  {journal} {\bibinfo  {journal} {Phys. Rev. B}\ }\textbf {\bibinfo {volume} {96}},\ \bibinfo {pages} {035150} (\bibinfo {year} {2017})}\BibitemShut {NoStop}%
\bibitem [{\citenamefont {Liu}\ \emph {et~al.}(2018)\citenamefont {Liu}, \citenamefont {Gromov},\ and\ \citenamefont {Papi\ifmmode~\acute{c}\else \'{c}\fi{}}}]{Liu2018Geometric}%
  \BibitemOpen
  \bibfield  {author} {\bibinfo {author} {\bibfnamefont {Z.}~\bibnamefont {Liu}}, \bibinfo {author} {\bibfnamefont {A.}~\bibnamefont {Gromov}},\ and\ \bibinfo {author} {\bibfnamefont {Z.}~\bibnamefont {Papi\ifmmode~\acute{c}\else \'{c}\fi{}}},\ }\href {https://doi.org/10.1103/PhysRevB.98.155140} {\bibfield  {journal} {\bibinfo  {journal} {Phys. Rev. B}\ }\textbf {\bibinfo {volume} {98}},\ \bibinfo {pages} {155140} (\bibinfo {year} {2018})}\BibitemShut {NoStop}%
\bibitem [{\citenamefont {Balram}\ \emph {et~al.}(2022)\citenamefont {Balram}, \citenamefont {Liu}, \citenamefont {Gromov},\ and\ \citenamefont {Papi\ifmmode~\acute{c}\else \'{c}\fi{}}}]{Balram2022Very}%
  \BibitemOpen
  \bibfield  {author} {\bibinfo {author} {\bibfnamefont {A.~C.}\ \bibnamefont {Balram}}, \bibinfo {author} {\bibfnamefont {Z.}~\bibnamefont {Liu}}, \bibinfo {author} {\bibfnamefont {A.}~\bibnamefont {Gromov}},\ and\ \bibinfo {author} {\bibfnamefont {Z.}~\bibnamefont {Papi\ifmmode~\acute{c}\else \'{c}\fi{}}},\ }\href {https://doi.org/10.1103/PhysRevX.12.021008} {\bibfield  {journal} {\bibinfo  {journal} {Phys. Rev. X}\ }\textbf {\bibinfo {volume} {12}},\ \bibinfo {pages} {021008} (\bibinfo {year} {2022})}\BibitemShut {NoStop}%
\bibitem [{\citenamefont {Parameswaran}\ \emph {et~al.}(2013)\citenamefont {Parameswaran}, \citenamefont {Roy},\ and\ \citenamefont {Sondhi}}]{Parameswaran_2013}%
  \BibitemOpen
  \bibfield  {author} {\bibinfo {author} {\bibfnamefont {S.~A.}\ \bibnamefont {Parameswaran}}, \bibinfo {author} {\bibfnamefont {R.}~\bibnamefont {Roy}},\ and\ \bibinfo {author} {\bibfnamefont {S.~L.}\ \bibnamefont {Sondhi}},\ }\href {https://doi.org/10.1016/j.crhy.2013.04.003} {\bibfield  {journal} {\bibinfo  {journal} {Comptes Rendus. Physique}\ }\textbf {\bibinfo {volume} {14}},\ \bibinfo {pages} {816–839} (\bibinfo {year} {2013})}\BibitemShut {NoStop}%
\bibitem [{\citenamefont {Sheng}\ \emph {et~al.}(2011)\citenamefont {Sheng}, \citenamefont {Gu}, \citenamefont {Sun},\ and\ \citenamefont {Sheng}}]{Sheng2011}%
  \BibitemOpen
  \bibfield  {author} {\bibinfo {author} {\bibfnamefont {D.}~\bibnamefont {Sheng}}, \bibinfo {author} {\bibfnamefont {Z.-C.}\ \bibnamefont {Gu}}, \bibinfo {author} {\bibfnamefont {K.}~\bibnamefont {Sun}},\ and\ \bibinfo {author} {\bibfnamefont {L.}~\bibnamefont {Sheng}},\ }\bibfield  {journal} {\bibinfo  {journal} {Nature Communications}\ }\textbf {\bibinfo {volume} {2}},\ \href {https://doi.org/10.1038/ncomms1380} {10.1038/ncomms1380} (\bibinfo {year} {2011})\BibitemShut {NoStop}%
\bibitem [{\citenamefont {Wang}\ \emph {et~al.}(2011)\citenamefont {Wang}, \citenamefont {Gu}, \citenamefont {Gong},\ and\ \citenamefont {Sheng}}]{Wang2011}%
  \BibitemOpen
  \bibfield  {author} {\bibinfo {author} {\bibfnamefont {Y.-F.}\ \bibnamefont {Wang}}, \bibinfo {author} {\bibfnamefont {Z.-C.}\ \bibnamefont {Gu}}, \bibinfo {author} {\bibfnamefont {C.-D.}\ \bibnamefont {Gong}},\ and\ \bibinfo {author} {\bibfnamefont {D.~N.}\ \bibnamefont {Sheng}},\ }\href {https://doi.org/10.1103/PhysRevLett.107.146803} {\bibfield  {journal} {\bibinfo  {journal} {Phys. Rev. Lett.}\ }\textbf {\bibinfo {volume} {107}},\ \bibinfo {pages} {146803} (\bibinfo {year} {2011})}\BibitemShut {NoStop}%
\bibitem [{\citenamefont {Regnault}\ and\ \citenamefont {Bernevig}(2011)}]{Regnault2011}%
  \BibitemOpen
  \bibfield  {author} {\bibinfo {author} {\bibfnamefont {N.}~\bibnamefont {Regnault}}\ and\ \bibinfo {author} {\bibfnamefont {B.~A.}\ \bibnamefont {Bernevig}},\ }\href {https://doi.org/10.1103/PhysRevX.1.021014} {\bibfield  {journal} {\bibinfo  {journal} {Phys. Rev. X}\ }\textbf {\bibinfo {volume} {1}},\ \bibinfo {pages} {021014} (\bibinfo {year} {2011})}\BibitemShut {NoStop}%
\bibitem [{\citenamefont {Neupert}\ \emph {et~al.}(2011)\citenamefont {Neupert}, \citenamefont {Santos}, \citenamefont {Chamon},\ and\ \citenamefont {Mudry}}]{neupert2011fractional}%
  \BibitemOpen
  \bibfield  {author} {\bibinfo {author} {\bibfnamefont {T.}~\bibnamefont {Neupert}}, \bibinfo {author} {\bibfnamefont {L.}~\bibnamefont {Santos}}, \bibinfo {author} {\bibfnamefont {C.}~\bibnamefont {Chamon}},\ and\ \bibinfo {author} {\bibfnamefont {C.}~\bibnamefont {Mudry}},\ }\href {https://doi.org/10.1103/PhysRevLett.106.236804} {\bibfield  {journal} {\bibinfo  {journal} {Phys. Rev. Lett.}\ }\textbf {\bibinfo {volume} {106}},\ \bibinfo {pages} {236804} (\bibinfo {year} {2011})}\BibitemShut {NoStop}%
\bibitem [{\citenamefont {Tang}\ \emph {et~al.}(2011)\citenamefont {Tang}, \citenamefont {Mei},\ and\ \citenamefont {Wen}}]{Tang2011}%
  \BibitemOpen
  \bibfield  {author} {\bibinfo {author} {\bibfnamefont {E.}~\bibnamefont {Tang}}, \bibinfo {author} {\bibfnamefont {J.-W.}\ \bibnamefont {Mei}},\ and\ \bibinfo {author} {\bibfnamefont {X.-G.}\ \bibnamefont {Wen}},\ }\href {https://doi.org/10.1103/PhysRevLett.106.236802} {\bibfield  {journal} {\bibinfo  {journal} {Phys. Rev. Lett.}\ }\textbf {\bibinfo {volume} {106}},\ \bibinfo {pages} {236802} (\bibinfo {year} {2011})}\BibitemShut {NoStop}%
\bibitem [{\citenamefont {Cai}\ \emph {et~al.}(2023)\citenamefont {Cai}, \citenamefont {Anderson}, \citenamefont {Wang}, \citenamefont {Zhang}, \citenamefont {Liu}, \citenamefont {Holtzmann}, \citenamefont {Zhang}, \citenamefont {Fan}, \citenamefont {Taniguchi}, \citenamefont {Watanabe}, \citenamefont {Ran}, \citenamefont {Cao}, \citenamefont {Fu}, \citenamefont {Xiao}, \citenamefont {Yao},\ and\ \citenamefont {Xu}}]{caiSignature2023}%
  \BibitemOpen
  \bibfield  {author} {\bibinfo {author} {\bibfnamefont {J.}~\bibnamefont {Cai}}, \bibinfo {author} {\bibfnamefont {E.}~\bibnamefont {Anderson}}, \bibinfo {author} {\bibfnamefont {C.}~\bibnamefont {Wang}}, \bibinfo {author} {\bibfnamefont {X.}~\bibnamefont {Zhang}}, \bibinfo {author} {\bibfnamefont {X.}~\bibnamefont {Liu}}, \bibinfo {author} {\bibfnamefont {W.}~\bibnamefont {Holtzmann}}, \bibinfo {author} {\bibfnamefont {Y.}~\bibnamefont {Zhang}}, \bibinfo {author} {\bibfnamefont {F.}~\bibnamefont {Fan}}, \bibinfo {author} {\bibfnamefont {T.}~\bibnamefont {Taniguchi}}, \bibinfo {author} {\bibfnamefont {K.}~\bibnamefont {Watanabe}}, \bibinfo {author} {\bibfnamefont {Y.}~\bibnamefont {Ran}}, \bibinfo {author} {\bibfnamefont {T.}~\bibnamefont {Cao}}, \bibinfo {author} {\bibfnamefont {L.}~\bibnamefont {Fu}}, \bibinfo {author} {\bibfnamefont {D.}~\bibnamefont {Xiao}}, \bibinfo {author} {\bibfnamefont {W.}~\bibnamefont {Yao}},\ and\ \bibinfo {author} {\bibfnamefont {X.}~\bibnamefont {Xu}},\ }\href
  {https://doi.org/10.1038/s41586-023-06289-w} {\bibfield  {journal} {\bibinfo  {journal} {Nature}\ }\textbf {\bibinfo {volume} {622}},\ \bibinfo {pages} {63–68} (\bibinfo {year} {2023})}\BibitemShut {NoStop}%
\bibitem [{\citenamefont {Zeng}\ \emph {et~al.}(2023)\citenamefont {Zeng}, \citenamefont {Xia}, \citenamefont {Kang}, \citenamefont {Zhu}, \citenamefont {Kn\"uppel}, \citenamefont {Vaswani}, \citenamefont {Watanabe}, \citenamefont {Taniguchi}, \citenamefont {Mak},\ and\ \citenamefont {Shan}}]{zengThermodynamic2023}%
  \BibitemOpen
  \bibfield  {author} {\bibinfo {author} {\bibfnamefont {Y.}~\bibnamefont {Zeng}}, \bibinfo {author} {\bibfnamefont {Z.}~\bibnamefont {Xia}}, \bibinfo {author} {\bibfnamefont {K.}~\bibnamefont {Kang}}, \bibinfo {author} {\bibfnamefont {J.}~\bibnamefont {Zhu}}, \bibinfo {author} {\bibfnamefont {P.}~\bibnamefont {Kn\"uppel}}, \bibinfo {author} {\bibfnamefont {C.}~\bibnamefont {Vaswani}}, \bibinfo {author} {\bibfnamefont {K.}~\bibnamefont {Watanabe}}, \bibinfo {author} {\bibfnamefont {T.}~\bibnamefont {Taniguchi}}, \bibinfo {author} {\bibfnamefont {K.~F.}\ \bibnamefont {Mak}},\ and\ \bibinfo {author} {\bibfnamefont {J.}~\bibnamefont {Shan}},\ }\href {https://doi.org/10.1038/s41586-023-06452-3} {\bibfield  {journal} {\bibinfo  {journal} {Nature}\ }\textbf {\bibinfo {volume} {622}},\ \bibinfo {pages} {69 } (\bibinfo {year} {2023})}\BibitemShut {NoStop}%
\bibitem [{\citenamefont {Xu}\ \emph {et~al.}(2023)\citenamefont {Xu}, \citenamefont {Sun}, \citenamefont {Jia}, \citenamefont {Liu}, \citenamefont {Xu}, \citenamefont {Li}, \citenamefont {Gu}, \citenamefont {Watanabe}, \citenamefont {Taniguchi}, \citenamefont {Tong}, \citenamefont {Jia}, \citenamefont {Shi}, \citenamefont {Jiang}, \citenamefont {Zhang}, \citenamefont {Liu},\ and\ \citenamefont {Li}}]{xu2023_fci}%
  \BibitemOpen
  \bibfield  {author} {\bibinfo {author} {\bibfnamefont {F.}~\bibnamefont {Xu}}, \bibinfo {author} {\bibfnamefont {Z.}~\bibnamefont {Sun}}, \bibinfo {author} {\bibfnamefont {T.}~\bibnamefont {Jia}}, \bibinfo {author} {\bibfnamefont {C.}~\bibnamefont {Liu}}, \bibinfo {author} {\bibfnamefont {C.}~\bibnamefont {Xu}}, \bibinfo {author} {\bibfnamefont {C.}~\bibnamefont {Li}}, \bibinfo {author} {\bibfnamefont {Y.}~\bibnamefont {Gu}}, \bibinfo {author} {\bibfnamefont {K.}~\bibnamefont {Watanabe}}, \bibinfo {author} {\bibfnamefont {T.}~\bibnamefont {Taniguchi}}, \bibinfo {author} {\bibfnamefont {B.}~\bibnamefont {Tong}}, \bibinfo {author} {\bibfnamefont {J.}~\bibnamefont {Jia}}, \bibinfo {author} {\bibfnamefont {Z.}~\bibnamefont {Shi}}, \bibinfo {author} {\bibfnamefont {S.}~\bibnamefont {Jiang}}, \bibinfo {author} {\bibfnamefont {Y.}~\bibnamefont {Zhang}}, \bibinfo {author} {\bibfnamefont {X.}~\bibnamefont {Liu}},\ and\ \bibinfo {author} {\bibfnamefont {T.}~\bibnamefont {Li}},\ }\href
  {https://doi.org/10.1103/PhysRevX.13.031037} {\bibfield  {journal} {\bibinfo  {journal} {Phys. Rev. X}\ }\textbf {\bibinfo {volume} {13}},\ \bibinfo {pages} {031037} (\bibinfo {year} {2023})}\BibitemShut {NoStop}%
\bibitem [{\citenamefont {Park}\ \emph {et~al.}(2023)\citenamefont {Park}, \citenamefont {Cai}, \citenamefont {Anderson}, \citenamefont {Zhang}, \citenamefont {Zhu}, \citenamefont {Liu}, \citenamefont {Wang}, \citenamefont {Holtzmann}, \citenamefont {Hu}, \citenamefont {Liu}, \citenamefont {Taniguchi}, \citenamefont {Watanabe}, \citenamefont {Chu}, \citenamefont {Cao}, \citenamefont {Fu}, \citenamefont {Yao}, \citenamefont {Chang}, \citenamefont {Cobden}, \citenamefont {Xiao},\ and\ \citenamefont {Xu}}]{park2023_fqah}%
  \BibitemOpen
  \bibfield  {author} {\bibinfo {author} {\bibfnamefont {H.}~\bibnamefont {Park}}, \bibinfo {author} {\bibfnamefont {J.}~\bibnamefont {Cai}}, \bibinfo {author} {\bibfnamefont {E.}~\bibnamefont {Anderson}}, \bibinfo {author} {\bibfnamefont {Y.}~\bibnamefont {Zhang}}, \bibinfo {author} {\bibfnamefont {J.}~\bibnamefont {Zhu}}, \bibinfo {author} {\bibfnamefont {X.}~\bibnamefont {Liu}}, \bibinfo {author} {\bibfnamefont {C.}~\bibnamefont {Wang}}, \bibinfo {author} {\bibfnamefont {W.}~\bibnamefont {Holtzmann}}, \bibinfo {author} {\bibfnamefont {C.}~\bibnamefont {Hu}}, \bibinfo {author} {\bibfnamefont {Z.}~\bibnamefont {Liu}}, \bibinfo {author} {\bibfnamefont {T.}~\bibnamefont {Taniguchi}}, \bibinfo {author} {\bibfnamefont {K.}~\bibnamefont {Watanabe}}, \bibinfo {author} {\bibfnamefont {J.-H.}\ \bibnamefont {Chu}}, \bibinfo {author} {\bibfnamefont {T.}~\bibnamefont {Cao}}, \bibinfo {author} {\bibfnamefont {L.}~\bibnamefont {Fu}}, \bibinfo {author} {\bibfnamefont {W.}~\bibnamefont {Yao}}, \bibinfo {author}
  {\bibfnamefont {C.-Z.}\ \bibnamefont {Chang}}, \bibinfo {author} {\bibfnamefont {D.}~\bibnamefont {Cobden}}, \bibinfo {author} {\bibfnamefont {D.}~\bibnamefont {Xiao}},\ and\ \bibinfo {author} {\bibfnamefont {X.}~\bibnamefont {Xu}},\ }\href {https://doi.org/10.1038/s41586-023-06536-0} {\bibfield  {journal} {\bibinfo  {journal} {Nature}\ }\textbf {\bibinfo {volume} {622}},\ \bibinfo {pages} {74} (\bibinfo {year} {2023})}\BibitemShut {NoStop}%
\bibitem [{\citenamefont {Lu}\ \emph {et~al.}(2024)\citenamefont {Lu}, \citenamefont {Han}, \citenamefont {Yao}, \citenamefont {Reddy}, \citenamefont {Yang}, \citenamefont {Seo}, \citenamefont {Watanabe}, \citenamefont {Taniguchi}, \citenamefont {Fu},\ and\ \citenamefont {Ju}}]{ZLu2023fqh_graphene}%
  \BibitemOpen
  \bibfield  {author} {\bibinfo {author} {\bibfnamefont {Z.}~\bibnamefont {Lu}}, \bibinfo {author} {\bibfnamefont {T.}~\bibnamefont {Han}}, \bibinfo {author} {\bibfnamefont {Y.}~\bibnamefont {Yao}}, \bibinfo {author} {\bibfnamefont {A.~P.}\ \bibnamefont {Reddy}}, \bibinfo {author} {\bibfnamefont {J.}~\bibnamefont {Yang}}, \bibinfo {author} {\bibfnamefont {J.}~\bibnamefont {Seo}}, \bibinfo {author} {\bibfnamefont {K.}~\bibnamefont {Watanabe}}, \bibinfo {author} {\bibfnamefont {T.}~\bibnamefont {Taniguchi}}, \bibinfo {author} {\bibfnamefont {L.}~\bibnamefont {Fu}},\ and\ \bibinfo {author} {\bibfnamefont {L.}~\bibnamefont {Ju}},\ }\href {https://doi.org/10.1038/s41586-023-07010-7} {\bibfield  {journal} {\bibinfo  {journal} {Nature}\ }\textbf {\bibinfo {volume} {626}},\ \bibinfo {pages} {759} (\bibinfo {year} {2024})}\BibitemShut {NoStop}%
\bibitem [{\citenamefont {Haldane}\ and\ \citenamefont {Rezayi}(1985)}]{Haldane1985}%
  \BibitemOpen
  \bibfield  {author} {\bibinfo {author} {\bibfnamefont {F.~D.~M.}\ \bibnamefont {Haldane}}\ and\ \bibinfo {author} {\bibfnamefont {E.~H.}\ \bibnamefont {Rezayi}},\ }\href {https://doi.org/10.1103/physrevlett.54.237} {\bibfield  {journal} {\bibinfo  {journal} {Physical Review Letters}\ }\textbf {\bibinfo {volume} {54}},\ \bibinfo {pages} {237–240} (\bibinfo {year} {1985})}\BibitemShut {NoStop}%
\bibitem [{\citenamefont {Repellin}\ \emph {et~al.}(2014)\citenamefont {Repellin}, \citenamefont {Neupert}, \citenamefont {Papi\ifmmode~\acute{c}\else \'{c}\fi{}},\ and\ \citenamefont {Regnault}}]{Repellin_2014}%
  \BibitemOpen
  \bibfield  {author} {\bibinfo {author} {\bibfnamefont {C.}~\bibnamefont {Repellin}}, \bibinfo {author} {\bibfnamefont {T.}~\bibnamefont {Neupert}}, \bibinfo {author} {\bibfnamefont {Z.}~\bibnamefont {Papi\ifmmode~\acute{c}\else \'{c}\fi{}}},\ and\ \bibinfo {author} {\bibfnamefont {N.}~\bibnamefont {Regnault}},\ }\href {https://doi.org/10.1103/PhysRevB.90.045114} {\bibfield  {journal} {\bibinfo  {journal} {Phys. Rev. B}\ }\textbf {\bibinfo {volume} {90}},\ \bibinfo {pages} {045114} (\bibinfo {year} {2014})}\BibitemShut {NoStop}%
\bibitem [{\citenamefont {Liu}\ \emph {et~al.}(2024)\citenamefont {Liu}, \citenamefont {Zhao},\ and\ \citenamefont {Xiang}}]{Liu2024}%
  \BibitemOpen
  \bibfield  {author} {\bibinfo {author} {\bibfnamefont {Y.}~\bibnamefont {Liu}}, \bibinfo {author} {\bibfnamefont {T.}~\bibnamefont {Zhao}},\ and\ \bibinfo {author} {\bibfnamefont {T.}~\bibnamefont {Xiang}},\ }\href {https://doi.org/10.1103/PhysRevB.110.195137} {\bibfield  {journal} {\bibinfo  {journal} {Phys. Rev. B}\ }\textbf {\bibinfo {volume} {110}},\ \bibinfo {pages} {195137} (\bibinfo {year} {2024})}\BibitemShut {NoStop}%
\bibitem [{\citenamefont {{Kousa}}\ \emph {et~al.}(2025)\citenamefont {{Kousa}}, \citenamefont {{Morales-Dur{\'a}n}}, \citenamefont {{Wolf}}, \citenamefont {{Khalaf}},\ and\ \citenamefont {{MacDonald}}}]{Kousa2025Theory}%
  \BibitemOpen
  \bibfield  {author} {\bibinfo {author} {\bibfnamefont {B.~M.}\ \bibnamefont {{Kousa}}}, \bibinfo {author} {\bibfnamefont {N.}~\bibnamefont {{Morales-Dur{\'a}n}}}, \bibinfo {author} {\bibfnamefont {T.~M.~R.}\ \bibnamefont {{Wolf}}}, \bibinfo {author} {\bibfnamefont {E.}~\bibnamefont {{Khalaf}}},\ and\ \bibinfo {author} {\bibfnamefont {A.~H.}\ \bibnamefont {{MacDonald}}},\ }\href {https://doi.org/10.48550/arXiv.2502.17574} {\bibfield  {journal} {\bibinfo  {journal} {arXiv e-prints}\ ,\ \bibinfo {eid} {arXiv:2502.17574}} (\bibinfo {year} {2025})},\ \Eprint {https://arxiv.org/abs/2502.17574} {arXiv:2502.17574 [cond-mat.mes-hall]} \BibitemShut {NoStop}%
\bibitem [{\citenamefont {{Yu}}\ \emph {et~al.}(2024)\citenamefont {{Yu}}, \citenamefont {{Herzog-Arbeitman}}, \citenamefont {{Kwan}}, \citenamefont {{Regnault}},\ and\ \citenamefont {{Bernevig}}}]{yu2024moire}%
  \BibitemOpen
  \bibfield  {author} {\bibinfo {author} {\bibfnamefont {J.}~\bibnamefont {{Yu}}}, \bibinfo {author} {\bibfnamefont {J.}~\bibnamefont {{Herzog-Arbeitman}}}, \bibinfo {author} {\bibfnamefont {Y.~H.}\ \bibnamefont {{Kwan}}}, \bibinfo {author} {\bibfnamefont {N.}~\bibnamefont {{Regnault}}},\ and\ \bibinfo {author} {\bibfnamefont {B.~A.}\ \bibnamefont {{Bernevig}}},\ }\href {https://doi.org/10.48550/arXiv.2407.13770} {\bibfield  {journal} {\bibinfo  {journal} {arXiv e-prints}\ ,\ \bibinfo {eid} {arXiv:2407.13770}} (\bibinfo {year} {2024})},\ \Eprint {https://arxiv.org/abs/2407.13770} {arXiv:2407.13770 [cond-mat.str-el]} \BibitemShut {NoStop}%
\bibitem [{\citenamefont {Abouelkomsan}\ \emph {et~al.}(2024)\citenamefont {Abouelkomsan}, \citenamefont {Reddy}, \citenamefont {Fu},\ and\ \citenamefont {Bergholtz}}]{Abouelkomsan2024Band}%
  \BibitemOpen
  \bibfield  {author} {\bibinfo {author} {\bibfnamefont {A.}~\bibnamefont {Abouelkomsan}}, \bibinfo {author} {\bibfnamefont {A.~P.}\ \bibnamefont {Reddy}}, \bibinfo {author} {\bibfnamefont {L.}~\bibnamefont {Fu}},\ and\ \bibinfo {author} {\bibfnamefont {E.~J.}\ \bibnamefont {Bergholtz}},\ }\href {https://doi.org/10.1103/PhysRevB.109.L121107} {\bibfield  {journal} {\bibinfo  {journal} {Phys. Rev. B}\ }\textbf {\bibinfo {volume} {109}},\ \bibinfo {pages} {L121107} (\bibinfo {year} {2024})}\BibitemShut {NoStop}%
\bibitem [{sup()}]{suppl}%
  \BibitemOpen
  \href@noop {} {\bibinfo {title} {Sm}},\ \bibinfo {note} {in this supplemental material, we provide the details of DMRG and TDVP implementations, more data about the ground state phase diagram of our FCI model, focusing on the FCI-Solid I and Solid I-Solid II transitions, and the spectra of the charge neutral modes, single-particle mode along different paths other than those shown in the main text, evolution of CGM as the system size increases, and benchmark with exact diagonalization.}\BibitemShut {Stop}%
\bibitem [{\citenamefont {{Lu}}\ \emph {et~al.}(2024)\citenamefont {{Lu}}, \citenamefont {{Wu}}, \citenamefont {{Chen}},\ and\ \citenamefont {{Meng}}}]{luVestigial2024}%
  \BibitemOpen
  \bibfield  {author} {\bibinfo {author} {\bibfnamefont {H.}~\bibnamefont {{Lu}}}, \bibinfo {author} {\bibfnamefont {H.-Q.}\ \bibnamefont {{Wu}}}, \bibinfo {author} {\bibfnamefont {B.-B.}\ \bibnamefont {{Chen}}},\ and\ \bibinfo {author} {\bibfnamefont {Z.~Y.}\ \bibnamefont {{Meng}}},\ }\href {https://doi.org/10.48550/arXiv.2408.07111} {\bibfield  {journal} {\bibinfo  {journal} {arXiv e-prints}\ ,\ \bibinfo {eid} {arXiv:2408.07111}} (\bibinfo {year} {2024})},\ \Eprint {https://arxiv.org/abs/2408.07111} {arXiv:2408.07111 [cond-mat.mes-hall]} \BibitemShut {NoStop}%
\bibitem [{\citenamefont {Haegeman}\ \emph {et~al.}(2011)\citenamefont {Haegeman}, \citenamefont {Cirac}, \citenamefont {Osborne}, \citenamefont {Pi\ifmmode~\check{z}\else \v{z}\fi{}orn}, \citenamefont {Verschelde},\ and\ \citenamefont {Verstraete}}]{Haegeman2011Time}%
  \BibitemOpen
  \bibfield  {author} {\bibinfo {author} {\bibfnamefont {J.}~\bibnamefont {Haegeman}}, \bibinfo {author} {\bibfnamefont {J.~I.}\ \bibnamefont {Cirac}}, \bibinfo {author} {\bibfnamefont {T.~J.}\ \bibnamefont {Osborne}}, \bibinfo {author} {\bibfnamefont {I.}~\bibnamefont {Pi\ifmmode~\check{z}\else \v{z}\fi{}orn}}, \bibinfo {author} {\bibfnamefont {H.}~\bibnamefont {Verschelde}},\ and\ \bibinfo {author} {\bibfnamefont {F.}~\bibnamefont {Verstraete}},\ }\href {https://doi.org/10.1103/PhysRevLett.107.070601} {\bibfield  {journal} {\bibinfo  {journal} {Phys. Rev. Lett.}\ }\textbf {\bibinfo {volume} {107}},\ \bibinfo {pages} {070601} (\bibinfo {year} {2011})}\BibitemShut {NoStop}%
\bibitem [{\citenamefont {Haegeman}\ \emph {et~al.}(2016)\citenamefont {Haegeman}, \citenamefont {Lubich}, \citenamefont {Oseledets}, \citenamefont {Vandereycken},\ and\ \citenamefont {Verstraete}}]{Haegeman2016Unifying}%
  \BibitemOpen
  \bibfield  {author} {\bibinfo {author} {\bibfnamefont {J.}~\bibnamefont {Haegeman}}, \bibinfo {author} {\bibfnamefont {C.}~\bibnamefont {Lubich}}, \bibinfo {author} {\bibfnamefont {I.}~\bibnamefont {Oseledets}}, \bibinfo {author} {\bibfnamefont {B.}~\bibnamefont {Vandereycken}},\ and\ \bibinfo {author} {\bibfnamefont {F.}~\bibnamefont {Verstraete}},\ }\href {https://doi.org/10.1103/PhysRevB.94.165116} {\bibfield  {journal} {\bibinfo  {journal} {Phys. Rev. B}\ }\textbf {\bibinfo {volume} {94}},\ \bibinfo {pages} {165116} (\bibinfo {year} {2016})}\BibitemShut {NoStop}%
\bibitem [{\citenamefont {White}(1992)}]{White1992Density}%
  \BibitemOpen
  \bibfield  {author} {\bibinfo {author} {\bibfnamefont {S.~R.}\ \bibnamefont {White}},\ }\href {https://doi.org/10.1103/PhysRevLett.69.2863} {\bibfield  {journal} {\bibinfo  {journal} {Phys. Rev. Lett.}\ }\textbf {\bibinfo {volume} {69}},\ \bibinfo {pages} {2863} (\bibinfo {year} {1992})}\BibitemShut {NoStop}%
\bibitem [{\citenamefont {Grushin}\ \emph {et~al.}(2015)\citenamefont {Grushin}, \citenamefont {Motruk}, \citenamefont {Zaletel},\ and\ \citenamefont {Pollmann}}]{Grushin2015Characterization}%
  \BibitemOpen
  \bibfield  {author} {\bibinfo {author} {\bibfnamefont {A.~G.}\ \bibnamefont {Grushin}}, \bibinfo {author} {\bibfnamefont {J.}~\bibnamefont {Motruk}}, \bibinfo {author} {\bibfnamefont {M.~P.}\ \bibnamefont {Zaletel}},\ and\ \bibinfo {author} {\bibfnamefont {F.}~\bibnamefont {Pollmann}},\ }\href {https://doi.org/10.1103/PhysRevB.91.035136} {\bibfield  {journal} {\bibinfo  {journal} {Phys. Rev. B}\ }\textbf {\bibinfo {volume} {91}},\ \bibinfo {pages} {035136} (\bibinfo {year} {2015})}\BibitemShut {NoStop}%
\bibitem [{\citenamefont {Luo}\ \emph {et~al.}(2020)\citenamefont {Luo}, \citenamefont {He}, \citenamefont {Zhou}, \citenamefont {Wang},\ and\ \citenamefont {Gong}}]{Luo_2020}%
  \BibitemOpen
  \bibfield  {author} {\bibinfo {author} {\bibfnamefont {W.-W.}\ \bibnamefont {Luo}}, \bibinfo {author} {\bibfnamefont {A.-L.}\ \bibnamefont {He}}, \bibinfo {author} {\bibfnamefont {Y.}~\bibnamefont {Zhou}}, \bibinfo {author} {\bibfnamefont {Y.-F.}\ \bibnamefont {Wang}},\ and\ \bibinfo {author} {\bibfnamefont {C.-D.}\ \bibnamefont {Gong}},\ }\href {https://doi.org/10.1103/PhysRevB.102.155120} {\bibfield  {journal} {\bibinfo  {journal} {Phys. Rev. B}\ }\textbf {\bibinfo {volume} {102}},\ \bibinfo {pages} {155120} (\bibinfo {year} {2020})}\BibitemShut {NoStop}%
\bibitem [{\citenamefont {Léonard}\ \emph {et~al.}(2023)\citenamefont {Léonard}, \citenamefont {Kim}, \citenamefont {Kwan}, \citenamefont {Segura}, \citenamefont {Grusdt}, \citenamefont {Repellin}, \citenamefont {Goldman},\ and\ \citenamefont {Greiner}}]{L_onard_2023}%
  \BibitemOpen
  \bibfield  {author} {\bibinfo {author} {\bibfnamefont {J.}~\bibnamefont {Léonard}}, \bibinfo {author} {\bibfnamefont {S.}~\bibnamefont {Kim}}, \bibinfo {author} {\bibfnamefont {J.}~\bibnamefont {Kwan}}, \bibinfo {author} {\bibfnamefont {P.}~\bibnamefont {Segura}}, \bibinfo {author} {\bibfnamefont {F.}~\bibnamefont {Grusdt}}, \bibinfo {author} {\bibfnamefont {C.}~\bibnamefont {Repellin}}, \bibinfo {author} {\bibfnamefont {N.}~\bibnamefont {Goldman}},\ and\ \bibinfo {author} {\bibfnamefont {M.}~\bibnamefont {Greiner}},\ }\href {https://doi.org/10.1038/s41586-023-06122-4} {\bibfield  {journal} {\bibinfo  {journal} {Nature}\ }\textbf {\bibinfo {volume} {619}},\ \bibinfo {pages} {495–499} (\bibinfo {year} {2023})}\BibitemShut {NoStop}%
\bibitem [{\citenamefont {Nie}\ \emph {et~al.}(2017)\citenamefont {Nie}, \citenamefont {Maharaj}, \citenamefont {Fradkin},\ and\ \citenamefont {Kivelson}}]{Nie2017vestigial}%
  \BibitemOpen
  \bibfield  {author} {\bibinfo {author} {\bibfnamefont {L.}~\bibnamefont {Nie}}, \bibinfo {author} {\bibfnamefont {A.~V.}\ \bibnamefont {Maharaj}}, \bibinfo {author} {\bibfnamefont {E.}~\bibnamefont {Fradkin}},\ and\ \bibinfo {author} {\bibfnamefont {S.~A.}\ \bibnamefont {Kivelson}},\ }\href {https://doi.org/10.1103/PhysRevB.96.085142} {\bibfield  {journal} {\bibinfo  {journal} {Phys. Rev. B}\ }\textbf {\bibinfo {volume} {96}},\ \bibinfo {pages} {085142} (\bibinfo {year} {2017})}\BibitemShut {NoStop}%
\bibitem [{\citenamefont {Fernandes}\ \emph {et~al.}(2019)\citenamefont {Fernandes}, \citenamefont {Orth},\ and\ \citenamefont {Schmalian}}]{Fernandes2019Vestigial}%
  \BibitemOpen
  \bibfield  {author} {\bibinfo {author} {\bibfnamefont {R.~M.}\ \bibnamefont {Fernandes}}, \bibinfo {author} {\bibfnamefont {P.~P.}\ \bibnamefont {Orth}},\ and\ \bibinfo {author} {\bibfnamefont {J.}~\bibnamefont {Schmalian}},\ }\href {https://doi.org/https://doi.org/10.1146/annurev-conmatphys-031218-013200} {\bibfield  {journal} {\bibinfo  {journal} {Annual Review of Condensed Matter Physics}\ }\textbf {\bibinfo {volume} {10}},\ \bibinfo {pages} {133} (\bibinfo {year} {2019})}\BibitemShut {NoStop}%
\bibitem [{\citenamefont {Wang}\ \emph {et~al.}(2021)\citenamefont {Wang}, \citenamefont {Yan}, \citenamefont {Wang}, \citenamefont {Qi},\ and\ \citenamefont {Meng}}]{wangVesigial2021}%
  \BibitemOpen
  \bibfield  {author} {\bibinfo {author} {\bibfnamefont {Y.-C.}\ \bibnamefont {Wang}}, \bibinfo {author} {\bibfnamefont {Z.}~\bibnamefont {Yan}}, \bibinfo {author} {\bibfnamefont {C.}~\bibnamefont {Wang}}, \bibinfo {author} {\bibfnamefont {Y.}~\bibnamefont {Qi}},\ and\ \bibinfo {author} {\bibfnamefont {Z.~Y.}\ \bibnamefont {Meng}},\ }\href {https://doi.org/10.1103/PhysRevB.103.014408} {\bibfield  {journal} {\bibinfo  {journal} {Phys. Rev. B}\ }\textbf {\bibinfo {volume} {103}},\ \bibinfo {pages} {014408} (\bibinfo {year} {2021})}\BibitemShut {NoStop}%
\bibitem [{\citenamefont {{Sun}}\ \emph {et~al.}(2024)\citenamefont {{Sun}}, \citenamefont {{Ye}}, \citenamefont {{Huang}}, \citenamefont {{Zhou}}, \citenamefont {{Huang}}, \citenamefont {{Li}}, \citenamefont {{Ye}}, \citenamefont {{Nnokwe}}, \citenamefont {{Deng}}, \citenamefont {{Mandrus}}, \citenamefont {{Meng}}, \citenamefont {{Sun}}, \citenamefont {{Du}}, \citenamefont {{He}},\ and\ \citenamefont {{Zhao}}}]{sun2024vestigial}%
  \BibitemOpen
  \bibfield  {author} {\bibinfo {author} {\bibfnamefont {Z.}~\bibnamefont {{Sun}}}, \bibinfo {author} {\bibfnamefont {G.}~\bibnamefont {{Ye}}}, \bibinfo {author} {\bibfnamefont {M.}~\bibnamefont {{Huang}}}, \bibinfo {author} {\bibfnamefont {C.}~\bibnamefont {{Zhou}}}, \bibinfo {author} {\bibfnamefont {N.}~\bibnamefont {{Huang}}}, \bibinfo {author} {\bibfnamefont {Q.}~\bibnamefont {{Li}}}, \bibinfo {author} {\bibfnamefont {Z.}~\bibnamefont {{Ye}}}, \bibinfo {author} {\bibfnamefont {C.}~\bibnamefont {{Nnokwe}}}, \bibinfo {author} {\bibfnamefont {H.}~\bibnamefont {{Deng}}}, \bibinfo {author} {\bibfnamefont {D.}~\bibnamefont {{Mandrus}}}, \bibinfo {author} {\bibfnamefont {Z.~Y.}\ \bibnamefont {{Meng}}}, \bibinfo {author} {\bibfnamefont {K.}~\bibnamefont {{Sun}}}, \bibinfo {author} {\bibfnamefont {C.}~\bibnamefont {{Du}}}, \bibinfo {author} {\bibfnamefont {R.}~\bibnamefont {{He}}},\ and\ \bibinfo {author} {\bibfnamefont {L.}~\bibnamefont {{Zhao}}},\ }\href {https://doi.org/10.1038/s41567-024-02618-6} {\bibfield
  {journal} {\bibinfo  {journal} {Nature Physics}\ }\textbf {\bibinfo {volume} {20}},\ \bibinfo {pages} {1764 } (\bibinfo {year} {2024})}\BibitemShut {NoStop}%
\bibitem [{\citenamefont {Svistunov}\ \emph {et~al.}(2015)\citenamefont {Svistunov}, \citenamefont {Babaev},\ and\ \citenamefont {Prokof'ev}}]{svistunov2015superfluid}%
  \BibitemOpen
  \bibfield  {author} {\bibinfo {author} {\bibfnamefont {B.~V.}\ \bibnamefont {Svistunov}}, \bibinfo {author} {\bibfnamefont {E.~S.}\ \bibnamefont {Babaev}},\ and\ \bibinfo {author} {\bibfnamefont {N.~V.}\ \bibnamefont {Prokof'ev}},\ }\href@noop {} {\emph {\bibinfo {title} {Superfluid states of matter}}}\ (\bibinfo  {publisher} {Crc Press},\ \bibinfo {year} {2015})\BibitemShut {NoStop}%
\bibitem [{\citenamefont {Grinenko}\ \emph {et~al.}(2021)\citenamefont {Grinenko}, \citenamefont {Weston}, \citenamefont {Caglieris}, \citenamefont {Wuttke}, \citenamefont {Hess}, \citenamefont {Gottschall}, \citenamefont {Maccari}, \citenamefont {Gorbunov}, \citenamefont {Zherlitsyn}, \citenamefont {Wosnitza} \emph {et~al.}}]{grinenko2021state}%
  \BibitemOpen
  \bibfield  {author} {\bibinfo {author} {\bibfnamefont {V.}~\bibnamefont {Grinenko}}, \bibinfo {author} {\bibfnamefont {D.}~\bibnamefont {Weston}}, \bibinfo {author} {\bibfnamefont {F.}~\bibnamefont {Caglieris}}, \bibinfo {author} {\bibfnamefont {C.}~\bibnamefont {Wuttke}}, \bibinfo {author} {\bibfnamefont {C.}~\bibnamefont {Hess}}, \bibinfo {author} {\bibfnamefont {T.}~\bibnamefont {Gottschall}}, \bibinfo {author} {\bibfnamefont {I.}~\bibnamefont {Maccari}}, \bibinfo {author} {\bibfnamefont {D.}~\bibnamefont {Gorbunov}}, \bibinfo {author} {\bibfnamefont {S.}~\bibnamefont {Zherlitsyn}}, \bibinfo {author} {\bibfnamefont {J.}~\bibnamefont {Wosnitza}}, \emph {et~al.},\ }\href@noop {} {\bibfield  {journal} {\bibinfo  {journal} {Nature Physics}\ }\textbf {\bibinfo {volume} {17}},\ \bibinfo {pages} {1254} (\bibinfo {year} {2021})}\BibitemShut {NoStop}%
\bibitem [{\citenamefont {Lu}\ \emph {et~al.}(2024)\citenamefont {Lu}, \citenamefont {Chen}, \citenamefont {Wu}, \citenamefont {Sun},\ and\ \citenamefont {Meng}}]{Lu2024Thermodynamics}%
  \BibitemOpen
  \bibfield  {author} {\bibinfo {author} {\bibfnamefont {H.}~\bibnamefont {Lu}}, \bibinfo {author} {\bibfnamefont {B.-B.}\ \bibnamefont {Chen}}, \bibinfo {author} {\bibfnamefont {H.-Q.}\ \bibnamefont {Wu}}, \bibinfo {author} {\bibfnamefont {K.}~\bibnamefont {Sun}},\ and\ \bibinfo {author} {\bibfnamefont {Z.~Y.}\ \bibnamefont {Meng}},\ }\href {https://doi.org/10.1103/PhysRevLett.132.236502} {\bibfield  {journal} {\bibinfo  {journal} {Phys. Rev. Lett.}\ }\textbf {\bibinfo {volume} {132}},\ \bibinfo {pages} {236502} (\bibinfo {year} {2024})}\BibitemShut {NoStop}%
\bibitem [{\citenamefont {{Lu}}\ \emph {et~al.}(2024)\citenamefont {{Lu}}, \citenamefont {{Wu}}, \citenamefont {{Chen}}, \citenamefont {{Sun}},\ and\ \citenamefont {{Meng}}}]{Lu2024Interaction}%
  \BibitemOpen
  \bibfield  {author} {\bibinfo {author} {\bibfnamefont {H.}~\bibnamefont {{Lu}}}, \bibinfo {author} {\bibfnamefont {H.-Q.}\ \bibnamefont {{Wu}}}, \bibinfo {author} {\bibfnamefont {B.-B.}\ \bibnamefont {{Chen}}}, \bibinfo {author} {\bibfnamefont {K.}~\bibnamefont {{Sun}}},\ and\ \bibinfo {author} {\bibfnamefont {Z.~Y.}\ \bibnamefont {{Meng}}},\ }\href {https://doi.org/10.48550/arXiv.2403.03258} {\bibfield  {journal} {\bibinfo  {journal} {arXiv e-prints}\ ,\ \bibinfo {eid} {arXiv:2403.03258}} (\bibinfo {year} {2024})},\ \Eprint {https://arxiv.org/abs/2403.03258} {arXiv:2403.03258 [cond-mat.str-el]} \BibitemShut {NoStop}%
\bibitem [{\citenamefont {Lu}\ \emph {et~al.}(2024)\citenamefont {Lu}, \citenamefont {Wu}, \citenamefont {Chen}, \citenamefont {Sun},\ and\ \citenamefont {Meng}}]{LuFractional2024}%
  \BibitemOpen
  \bibfield  {author} {\bibinfo {author} {\bibfnamefont {H.}~\bibnamefont {Lu}}, \bibinfo {author} {\bibfnamefont {H.-Q.}\ \bibnamefont {Wu}}, \bibinfo {author} {\bibfnamefont {B.-B.}\ \bibnamefont {Chen}}, \bibinfo {author} {\bibfnamefont {K.}~\bibnamefont {Sun}},\ and\ \bibinfo {author} {\bibfnamefont {Z.~Y.}\ \bibnamefont {Meng}},\ }\href {https://doi.org/10.1088/1361-6633/ad7640} {\bibfield  {journal} {\bibinfo  {journal} {Reports on Progress in Physics}\ }\textbf {\bibinfo {volume} {87}},\ \bibinfo {pages} {108003} (\bibinfo {year} {2024})}\BibitemShut {NoStop}%
\bibitem [{\citenamefont {{Lu}}\ \emph {et~al.}(2024)\citenamefont {{Lu}}, \citenamefont {{Wu}}, \citenamefont {{Chen}},\ and\ \citenamefont {{Meng}}}]{luFrom2024}%
  \BibitemOpen
  \bibfield  {author} {\bibinfo {author} {\bibfnamefont {H.}~\bibnamefont {{Lu}}}, \bibinfo {author} {\bibfnamefont {H.-Q.}\ \bibnamefont {{Wu}}}, \bibinfo {author} {\bibfnamefont {B.-B.}\ \bibnamefont {{Chen}}},\ and\ \bibinfo {author} {\bibfnamefont {Z.~Y.}\ \bibnamefont {{Meng}}},\ }\href {https://doi.org/10.48550/arXiv.2404.06745} {\bibfield  {journal} {\bibinfo  {journal} {arXiv e-prints}\ ,\ \bibinfo {eid} {arXiv:2404.06745}} (\bibinfo {year} {2024})},\ \Eprint {https://arxiv.org/abs/2404.06745} {arXiv:2404.06745 [cond-mat.str-el]} \BibitemShut {NoStop}%
\bibitem [{\citenamefont {Pan}\ \emph {et~al.}(2023)\citenamefont {Pan}, \citenamefont {Zhang}, \citenamefont {Lu}, \citenamefont {Li}, \citenamefont {Chen}, \citenamefont {Sun},\ and\ \citenamefont {Meng}}]{panThermodynamic2023}%
  \BibitemOpen
  \bibfield  {author} {\bibinfo {author} {\bibfnamefont {G.}~\bibnamefont {Pan}}, \bibinfo {author} {\bibfnamefont {X.}~\bibnamefont {Zhang}}, \bibinfo {author} {\bibfnamefont {H.}~\bibnamefont {Lu}}, \bibinfo {author} {\bibfnamefont {H.}~\bibnamefont {Li}}, \bibinfo {author} {\bibfnamefont {B.-B.}\ \bibnamefont {Chen}}, \bibinfo {author} {\bibfnamefont {K.}~\bibnamefont {Sun}},\ and\ \bibinfo {author} {\bibfnamefont {Z.~Y.}\ \bibnamefont {Meng}},\ }\href {https://doi.org/10.1103/PhysRevLett.130.016401} {\bibfield  {journal} {\bibinfo  {journal} {Phys. Rev. Lett.}\ }\textbf {\bibinfo {volume} {130}},\ \bibinfo {pages} {016401} (\bibinfo {year} {2023})}\BibitemShut {NoStop}%
\bibitem [{\citenamefont {Lin}\ \emph {et~al.}(2022)\citenamefont {Lin}, \citenamefont {Chen}, \citenamefont {Li}, \citenamefont {Meng},\ and\ \citenamefont {Shi}}]{excitonLin2022}%
  \BibitemOpen
  \bibfield  {author} {\bibinfo {author} {\bibfnamefont {X.}~\bibnamefont {Lin}}, \bibinfo {author} {\bibfnamefont {B.-B.}\ \bibnamefont {Chen}}, \bibinfo {author} {\bibfnamefont {W.}~\bibnamefont {Li}}, \bibinfo {author} {\bibfnamefont {Z.~Y.}\ \bibnamefont {Meng}},\ and\ \bibinfo {author} {\bibfnamefont {T.}~\bibnamefont {Shi}},\ }\href {https://doi.org/10.1103/PhysRevLett.128.157201} {\bibfield  {journal} {\bibinfo  {journal} {Phys. Rev. Lett.}\ }\textbf {\bibinfo {volume} {128}},\ \bibinfo {pages} {157201} (\bibinfo {year} {2022})}\BibitemShut {NoStop}%
\bibitem [{\citenamefont {{Lu}}\ \emph {et~al.}(2025)\citenamefont {{Lu}}, \citenamefont {{Wu}}, \citenamefont {{Chen}}, \citenamefont {{Yao}},\ and\ \citenamefont {{Meng}}}]{luGeneric2025}%
  \BibitemOpen
  \bibfield  {author} {\bibinfo {author} {\bibfnamefont {H.}~\bibnamefont {{Lu}}}, \bibinfo {author} {\bibfnamefont {H.-Q.}\ \bibnamefont {{Wu}}}, \bibinfo {author} {\bibfnamefont {B.-B.}\ \bibnamefont {{Chen}}}, \bibinfo {author} {\bibfnamefont {W.}~\bibnamefont {{Yao}}},\ and\ \bibinfo {author} {\bibfnamefont {Z.~Y.}\ \bibnamefont {{Meng}}},\ }\href {https://doi.org/10.48550/arXiv.2505.04138} {\bibfield  {journal} {\bibinfo  {journal} {arXiv e-prints}\ ,\ \bibinfo {eid} {arXiv:2505.04138}} (\bibinfo {year} {2025})},\ \Eprint {https://arxiv.org/abs/2505.04138} {arXiv:2505.04138 [cond-mat.str-el]} \BibitemShut {NoStop}%
\bibitem [{\citenamefont {Golkar}\ \emph {et~al.}(2016)\citenamefont {Golkar}, \citenamefont {Nguyen},\ and\ \citenamefont {Son}}]{Golkar2016}%
  \BibitemOpen
  \bibfield  {author} {\bibinfo {author} {\bibfnamefont {S.}~\bibnamefont {Golkar}}, \bibinfo {author} {\bibfnamefont {D.~X.}\ \bibnamefont {Nguyen}},\ and\ \bibinfo {author} {\bibfnamefont {D.~T.}\ \bibnamefont {Son}},\ }\bibfield  {journal} {\bibinfo  {journal} {Journal of High Energy Physics}\ }\textbf {\bibinfo {volume} {2016}},\ \href {https://doi.org/10.1007/jhep01(2016)021} {10.1007/jhep01(2016)021} (\bibinfo {year} {2016})\BibitemShut {NoStop}%
\bibitem [{\citenamefont {Yuzhu}\ and\ \citenamefont {Bo}(2023)}]{Yuzhu2023Geometric}%
  \BibitemOpen
  \bibfield  {author} {\bibinfo {author} {\bibfnamefont {W.}~\bibnamefont {Yuzhu}}\ and\ \bibinfo {author} {\bibfnamefont {Y.}~\bibnamefont {Bo}},\ }\bibfield  {journal} {\bibinfo  {journal} {Nature Communications}\ }\textbf {\bibinfo {volume} {14}},\ \href {https://doi.org/10.1038/s41467-023-38036-0} {10.1038/s41467-023-38036-0} (\bibinfo {year} {2023})\BibitemShut {NoStop}%
\bibitem [{\citenamefont {Wang}\ \emph {et~al.}(2025)\citenamefont {Wang}, \citenamefont {Huxford}, \citenamefont {Nguyen}, \citenamefont {Ji}, \citenamefont {Kim},\ and\ \citenamefont {Yang}}]{Yuzhu2025Dynamics}%
  \BibitemOpen
  \bibfield  {author} {\bibinfo {author} {\bibfnamefont {Y.}~\bibnamefont {Wang}}, \bibinfo {author} {\bibfnamefont {J.}~\bibnamefont {Huxford}}, \bibinfo {author} {\bibfnamefont {D.~X.}\ \bibnamefont {Nguyen}}, \bibinfo {author} {\bibfnamefont {G.}~\bibnamefont {Ji}}, \bibinfo {author} {\bibfnamefont {Y.~B.}\ \bibnamefont {Kim}},\ and\ \bibinfo {author} {\bibfnamefont {B.}~\bibnamefont {Yang}},\ }\href {https://doi.org/10.48550/ARXIV.2502.02640} {\bibinfo {title} {Dynamics and lifetime of geometric excitations in moiré systems}} (\bibinfo {year} {2025})\BibitemShut {NoStop}%
\bibitem [{\citenamefont {Hirjibehedin}\ \emph {et~al.}(2005)\citenamefont {Hirjibehedin}, \citenamefont {Dujovne}, \citenamefont {Pinczuk}, \citenamefont {Dennis}, \citenamefont {Pfeiffer},\ and\ \citenamefont {West}}]{hirjibehedinSplitting2005}%
  \BibitemOpen
  \bibfield  {author} {\bibinfo {author} {\bibfnamefont {C.~F.}\ \bibnamefont {Hirjibehedin}}, \bibinfo {author} {\bibfnamefont {I.}~\bibnamefont {Dujovne}}, \bibinfo {author} {\bibfnamefont {A.}~\bibnamefont {Pinczuk}}, \bibinfo {author} {\bibfnamefont {B.~S.}\ \bibnamefont {Dennis}}, \bibinfo {author} {\bibfnamefont {L.~N.}\ \bibnamefont {Pfeiffer}},\ and\ \bibinfo {author} {\bibfnamefont {K.~W.}\ \bibnamefont {West}},\ }\href {https://doi.org/10.1103/PhysRevLett.95.066803} {\bibfield  {journal} {\bibinfo  {journal} {Phys. Rev. Lett.}\ }\textbf {\bibinfo {volume} {95}},\ \bibinfo {pages} {066803} (\bibinfo {year} {2005})}\BibitemShut {NoStop}%
\bibitem [{\citenamefont {{Shen}}\ \emph {et~al.}(2024)\citenamefont {{Shen}}, \citenamefont {{Wang}}, \citenamefont {{Hu}}, \citenamefont {{Guo}}, \citenamefont {{Yao}}, \citenamefont {{Wang}}, \citenamefont {{Duan}},\ and\ \citenamefont {{Xu}}}]{shenMagnetorotons2024}%
  \BibitemOpen
  \bibfield  {author} {\bibinfo {author} {\bibfnamefont {X.}~\bibnamefont {{Shen}}}, \bibinfo {author} {\bibfnamefont {C.}~\bibnamefont {{Wang}}}, \bibinfo {author} {\bibfnamefont {X.}~\bibnamefont {{Hu}}}, \bibinfo {author} {\bibfnamefont {R.}~\bibnamefont {{Guo}}}, \bibinfo {author} {\bibfnamefont {H.}~\bibnamefont {{Yao}}}, \bibinfo {author} {\bibfnamefont {C.}~\bibnamefont {{Wang}}}, \bibinfo {author} {\bibfnamefont {W.}~\bibnamefont {{Duan}}},\ and\ \bibinfo {author} {\bibfnamefont {Y.}~\bibnamefont {{Xu}}},\ }\href {https://doi.org/10.48550/arXiv.2412.01211} {\bibfield  {journal} {\bibinfo  {journal} {arXiv e-prints}\ ,\ \bibinfo {eid} {arXiv:2412.01211}} (\bibinfo {year} {2024})},\ \Eprint {https://arxiv.org/abs/2412.01211} {arXiv:2412.01211 [cond-mat.str-el]} \BibitemShut {NoStop}%
\bibitem [{Ten()}]{Tensorkit_web}%
  \BibitemOpen
  \href@noop {} {\bibinfo {title} {Tensorkit}},\ \bibinfo {howpublished} {\url{https://jutho.github.io/TensorKit.jl/stable/}},\ \bibinfo {note} {accessed: December 21, 2024}\BibitemShut {NoStop}%
\end{thebibliography}%

\newpage\clearpage
\renewcommand{\theequation}{S\arabic{equation}} \renewcommand{\thefigure}{S%
	\arabic{figure}} \setcounter{equation}{0} \setcounter{figure}{0}

\begin{widetext}
	
In this supplemental material, we provide the details of DMRG and TDVP implementations, more data about the ground state phase diagram of our FCI model, focusing on the FCI-Solid I and Solid I-Solid II transitions, and the spectra of the charge neutral modes, single-particle mode along different paths other than those shown in the main text, evolution of CGM as the system size increases, and benchmark with exact diagonalization.  

\subsection{Section I: DMRG and TDVP implementation details}
\begin{figure}[htp!]
	\centering		
	\includegraphics[width=0.89\textwidth]{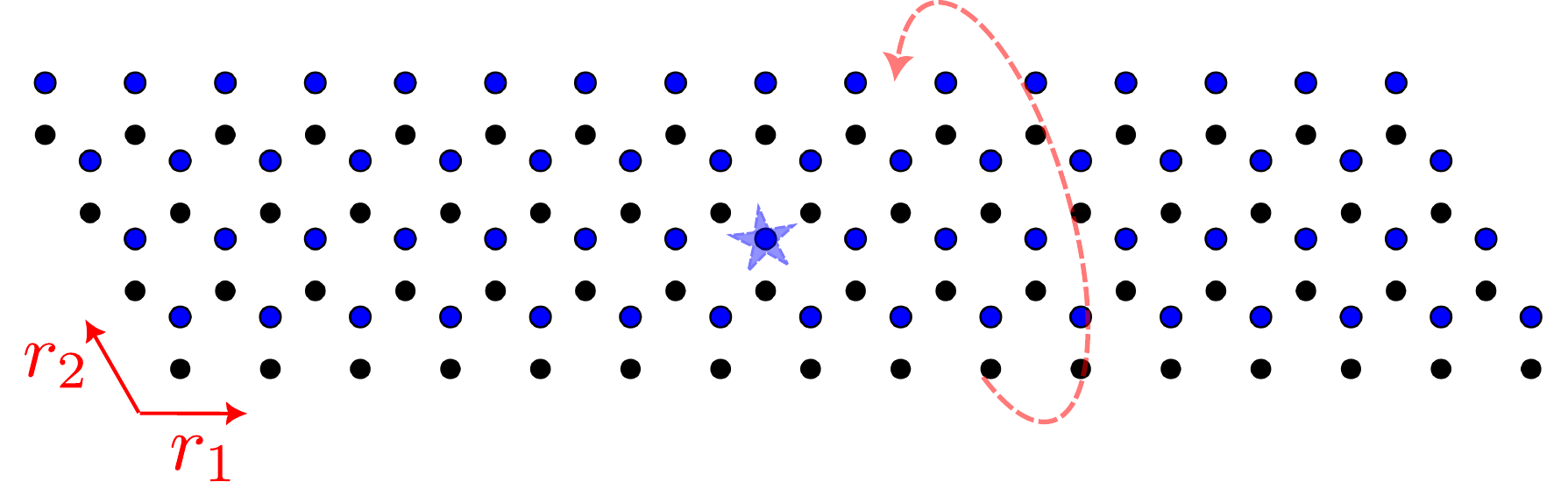}
	\caption{Configuration of $4\times 16$ cylinder with periodic boundary condition along the $\mathbf{r_2}$ direction. The start is the location of $\mathbf{r^{A}_0}$. }.  
	\label{fig_figS1}
\end{figure}	
In this section, we describe the configuration of the finite honeycomb lattice under investigation and the details of DMRG and TDVP simulations.

Fig.~\ref{fig_figS1} displays the $N_1 \times N_2 \times 2 = 16 \times 4  \times2$ cylinders that our calculation mainly based on. The black and blue dots label the $A$ and $B$ sub-lattices, respectively. We impose periodic boundary condition along $\mathbf{r_2}$ direction and open boundary condition along $\mathbf{r_1}$ direction. The blue star labels the position of $\mathbf{r^A_0}$ where we act an excitation (e.g. $n^A_0$) on ground state and time evolve the wave function. We use the same $\mathbf{r^A_0}$ for the calculation of  $S^A(\mathbf{k},\omega)$, $G^A(\mathbf{k},\omega)$ and $g^\pm(\mathbf{k},\omega)$.

Charge $U(1)$ symmetry is implemented in both DMRG and TDVP calculation based on the TensorKit package~\cite{Tensorkit_web}. We keep the bond dimension up to $m=2700$ states in the DMRG simulation to ensure the maximum truncation error below $10^{-5}$. In TDVP simulation, we keep up to $m=250$ states ensuring maximum truncation error below $6\times10^{-4}$, and time evolve wavefunction $N_t = 2000$ steps with $\Delta t = 0.05$, resulting energy resolution $\Delta \omega  = \frac{1}{N_t\Delta t}= 0.01$ upto $\max(\omega) = \frac{1}{\Delta t} = 20$, we set $\hbar = 1$ in time evolution.  

We discuss in detail the convergence of DMRG and TDVP simulation with bond dimension in the following section.

\subsection{Section II: Convergence of DMRG simulation and TDVP numerical stabilization}
In this section we show the convergence of the entanglement entropy $S_\mathrm{E}$ of the ground state in the FCI, Solid I and Solid II phases.

As shown in Fig.~\ref{fig_S_EE_dmrg} (a), the $S_E$ of the ground state wave function in the FCI phase saturates to constant value along the $\mathbf{r_1}$ direction as the bond dimension $D$ increases, which is consistent with the fact tat the topological ordered state is gapped in the bulk. Fig.~\ref{fig_S_EE_dmrg} (b) and (c) show the $S_E$ in the Solid I and Solid II phases, the oscillations we observed, resemble the charge order inside the CDW phases.


Since our TDVP simulation focuses on the FCI ground state with relatively small bond dimension. We also examine the overlap of the ground state wave function achieved by DMRG at different bond dimensions, the overlap of the ground state wave function is remarkably accurate even in small bond dimenstion with value $\langle \psi_{D = 200} | \psi_{D = 1200}\rangle \approx 99.9\%$ , which indicates that the MPS in small bond dimension already encodes the majority information of the FCI ground state and greatly facilitate our TDVP simulation. 

\begin{figure}[htp!]
	\centering		
	\includegraphics[width=0.89\textwidth]{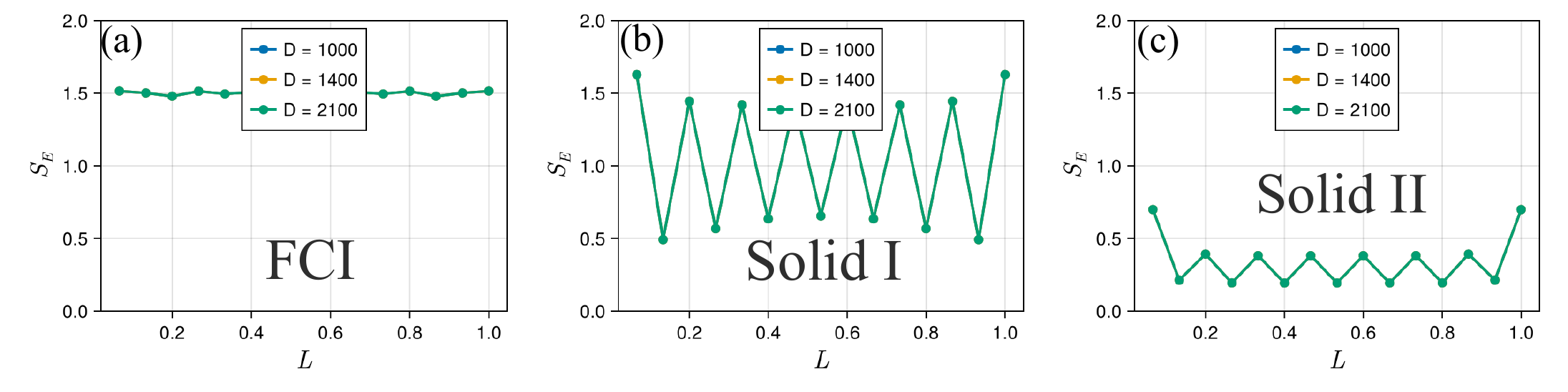}
	\caption{Entanglement entropy of FCI (a) with $V=1$, Solid I (b) with $V=3.6$ and Solid II (c) with $V=4.6$ phases. Here, $i$ measures the distance of the unit cell from the edge of the cylinder along $\mathbf{r_1}$ direction.} 
    \label{fig_S_EE_dmrg}
\end{figure}   

\begin{figure}[htp!]
	\centering		
	\includegraphics[width=0.89\textwidth]{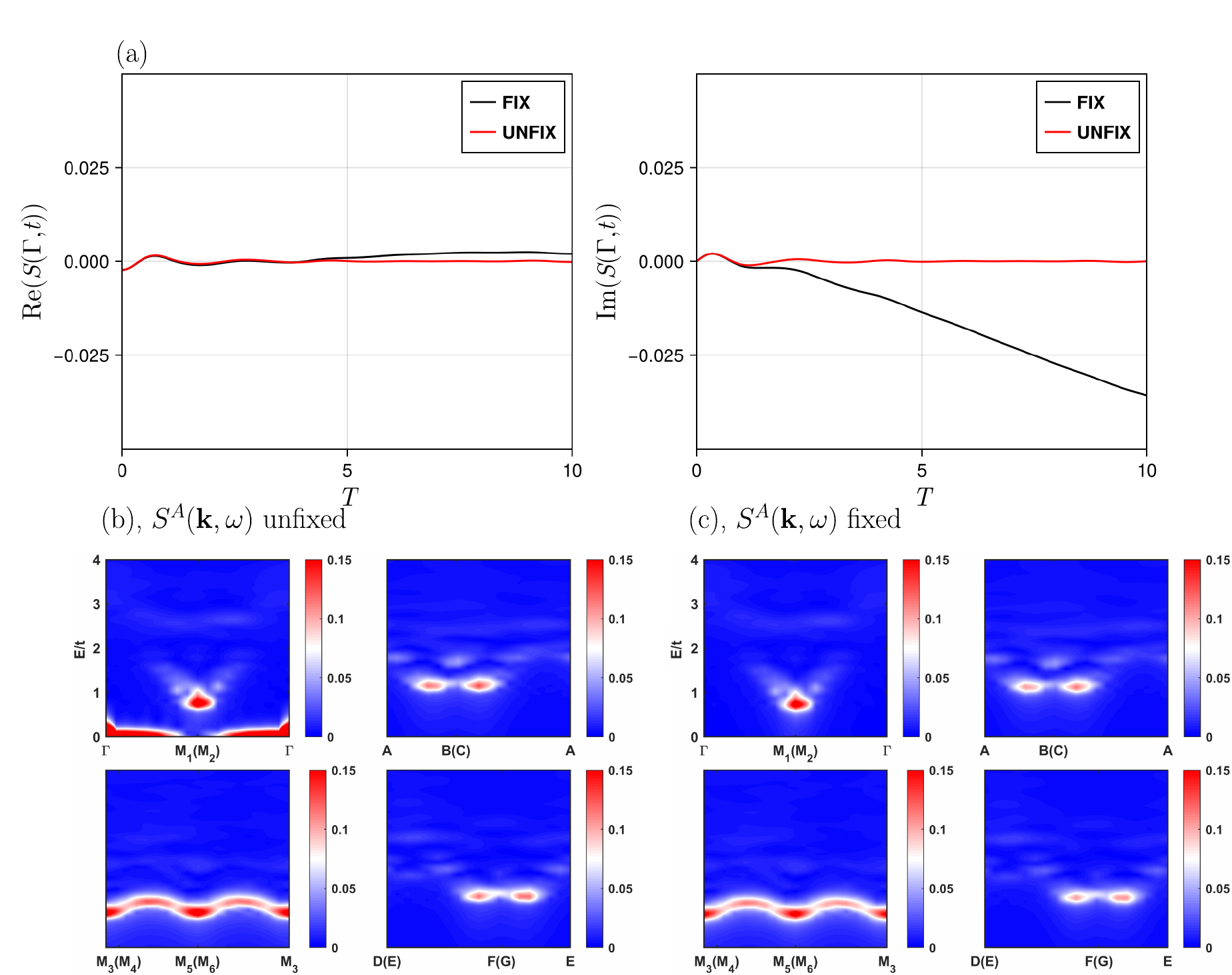}
	\caption{(a), (b) are the time-dependent dynamic density structure factor with a global factor unfixed or fixed, respectively. (c) and (d) represent the respective spectra of (a) and (b).  The calculation is done at $V_1 = 4, V_2 = 1$ and the momentum labeling is illustrated in Fig.~\ref{fig:fig1} (d) of the main text. }
    \label{fig_S_tdvp_D}
\end{figure}	

Next, we show the convergence of the TDVP calculation with respect to the bond dimension and length of cylinder. As well as the fixing scheme we adopt to ensure the stablity of the TDVP simulation.

\noindent{\textcolor{blue}{\it Bond dimension.}---} In Fig.~\ref{fig_S_tdvp_D} (a) and (b), we show the convergence of the dynamic density structure factor with respect to the bond dimension. While in Fig.~\ref{fig_S_tdvp_D} (a), the $S^{A}(\mathbf{k} = \mathbf{M_3},t)$ gradually exhibits converged behavior as the bond dimension increases. The  $S^{A}(\mathbf{k=\mathbf{\Gamma}},t)$  behaves differently upon varying bond dimension. Moreover, in Fig.~\ref{fig_S_tdvp_D} (c), $S^{A}(\mathbf{k = \Gamma},\omega = 0)$ present a sharp static peak which is unphysical, since the neutral mode at $\Gamma$ point represents a global shift of charge density which is contradictory to the conservation of charge. Therefore, in the calculation of $S^A(\mathbf{k},\omega)$ and $g^\pm(\mathbf{k},\omega)$, we rescale the canonical center of the matrix product states (MPS) after every TDVP sweep to ensure the amplitude of the wavefunction is conserved,
\begin{equation}
    \langle \psi_0 | \psi(t) \rangle = e^{-iE_0t} \langle \psi_0 |f(\hat{n}_{\{i\}})| \psi_0 \rangle
\end{equation}
where $|\psi_0\rangle$ is the ground state, $|\psi(t)\rangle$ is the time evolved state which initally is $f(\hat{n}_{\{i\}})|\psi_0\rangle$ at time $t$, $E_0$ is the ground state energy, $f(\hat{n}_{\{i\}})$ represent any combination of neutral excitation operator. The right hand side of the equation can be determined before the TDVP simulation, then be used to assist the stability of TDVP. We note that such fixed scheme not only fix the norm of the wave function but also the global phase, which is crucial for the determination of the energy gap.

The spectrum of the fixed and unfixed dynamic density structure factor are shown in Fig.~\ref{fig_S_tdvp_D} (c) and (d), respectively. The unphysical static peak at $\Gamma$ point is removed in the fixed scheme while the other gapped mode remains its' shape and position unchanged. This could also be seen from the comparison of $S^A(\mathbf{k = M_3},t)$ between Fig.~\ref{fig_S_tdvp_D} (a) and (b), where they almost have the same manner as time evolves . The first roton gap $\Delta_{n_1}$ that is invisible due to the sharp weight of static peak in Fig.~\ref{fig_S_tdvp_D} (c) can be clearly seen in the fixed scheme Fig.~\ref{fig_S_tdvp_D} (d). For the calculation of $G^A({\mathbf{k},\omega})$, we fix the norm of the wave function in the same way. 

\noindent{\textcolor{blue}{\it Length of cylinder.}---} We also increase the length ($N_1$) of the cylinder to ensure the convergence of the roton gaps $\Delta_{n_1}$ and  $\Delta_{n_2}$.  As shown in Fig.~\ref{fig_S_length_gap}, the roton gaps $\Delta_{n_1}$ and $\Delta_{n_2}$ are robust as the cylinder length increases, which is consistent with the fact that the roton mode is a bulk excitation in the FCI phase, and exclude the possibility of finite size and boundary effects in our calculation.

For the calculation of $g^\pm(\mathbf{k},\omega)$, we monitor the evolution of spectrum weight upon increasing the cylinder length, as shown in Fig.~\ref{fig_S_length_graviton}. We observe a reduction in spectral weight with increasing cylinder length. This trend is attributed to the increase of density of states at larger systems, introducing more scattering channels and consequently lowering the lifetime of CGM\cite{Yuzhu2025Dynamics}.
\begin{figure}[h!]
	\centering		
	\includegraphics[width=0.4\textwidth]{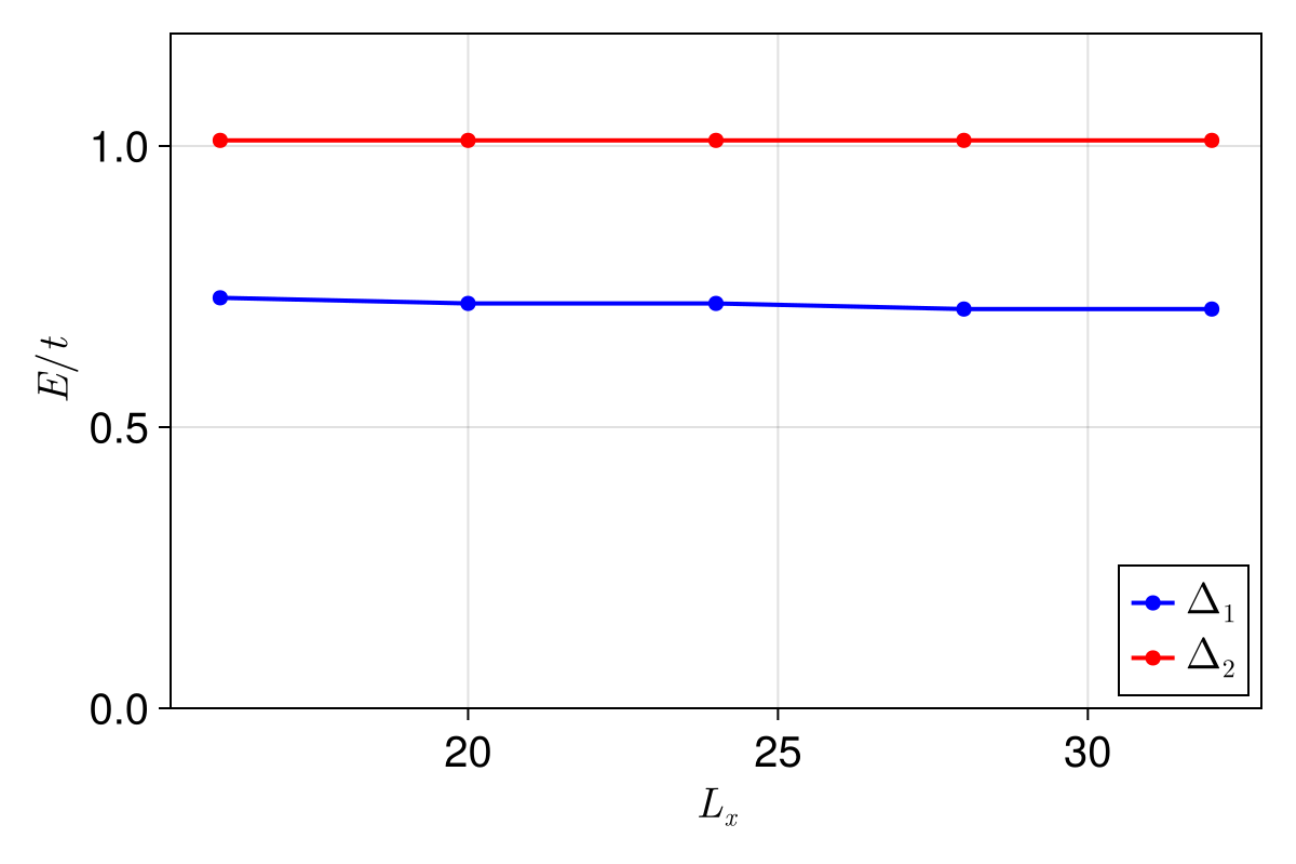}
	\caption{$V_1 = 4, V_2 = 1,D = 200$,  as the cylinder length increase. $\Delta_{n_1}$ and $\Delta_{n_2}$ are robust as the cylinder length increases. }
    \label{fig_S_length_gap}
\end{figure}	

\begin{figure}[h!]
	\centering		
	\includegraphics[width=0.4\textwidth]{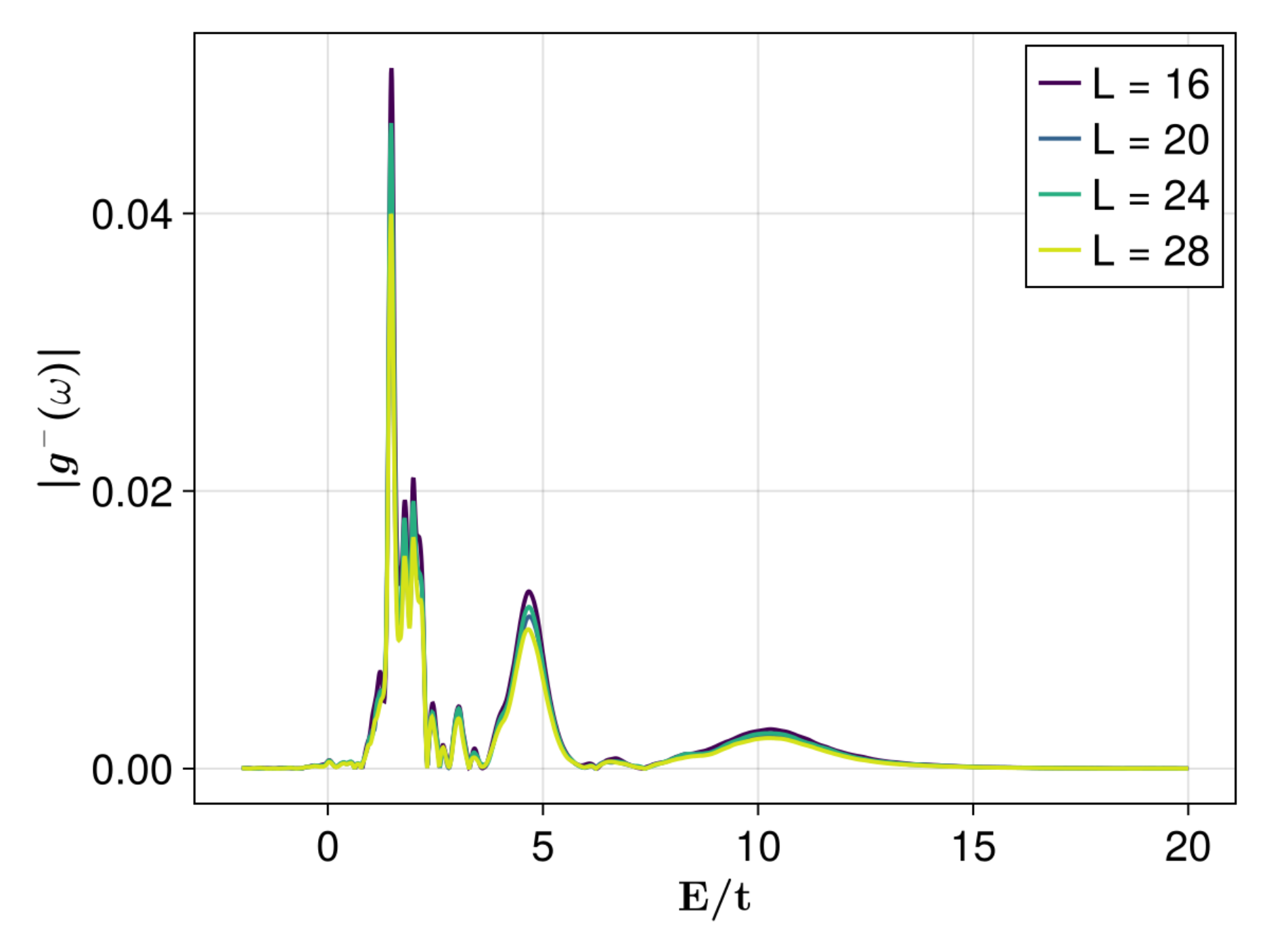}
	\caption{Evolution of CGM spectrum weight as the cylinder length increases at $V_1 = 4, V_2 = 1,D = 200, L_y = 4$.  Here we show the result at the $\Gamma$ point where the CGM is centered, the spectrum peak gradually declines as we increase the length of the cylinder.}
    \label{fig_S_length_graviton}
\end{figure}

\subsection{Section IV: FCI-Solid I and Solid I-Solid II transitions}
\begin{figure}[htp!]
	\centering		
	\includegraphics[width=0.89\textwidth]{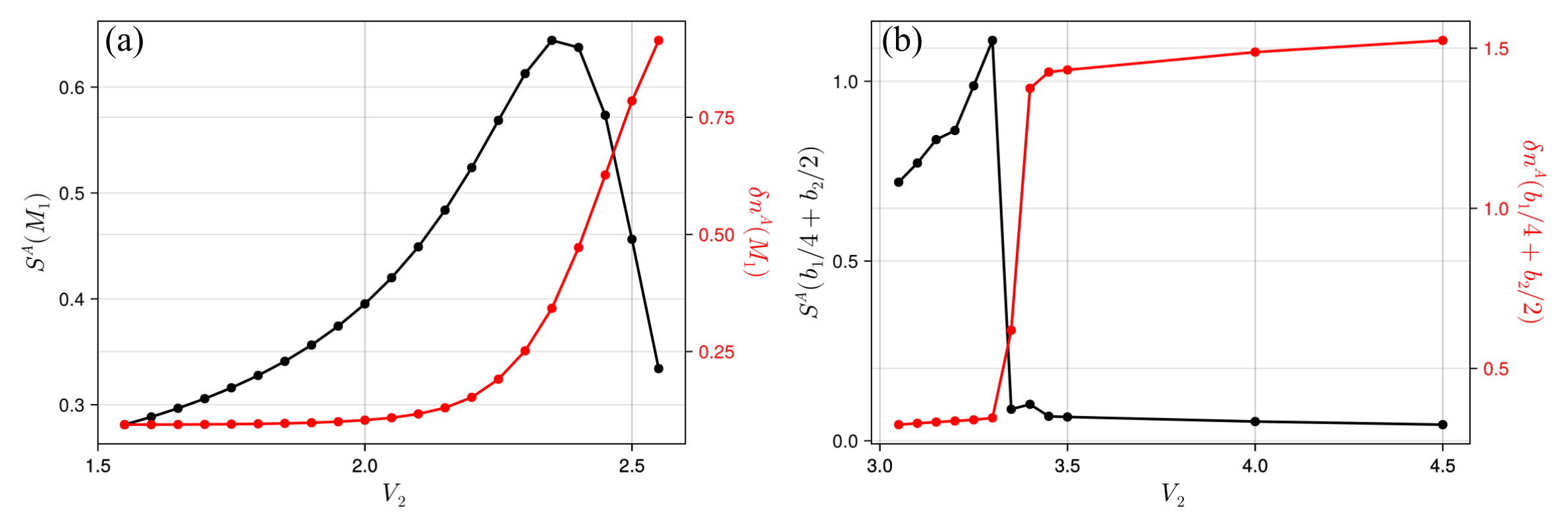}
	\caption{FCI-Solid I transition with charge order $\delta n^A(\mathbf{M_1})$ (a) and static density structure factor $S^A(\mathbf{M_1})$ (b). (c) and (d) are the $\delta n^A(\mathbf{M_1})$ and $S^A(\mathbf{k})$ measured at $\mathbf{b_1}/4 + \mathbf{b_2}/2.$} 
	\label{fig_S_phase_transition}
\end{figure}	
In this section, we determine the phase boundary between FCI, Solid I and Solid II based on DMRG calculation. 

\begin{figure}[h!]
        \centering
        \includegraphics[width=0.8\linewidth]{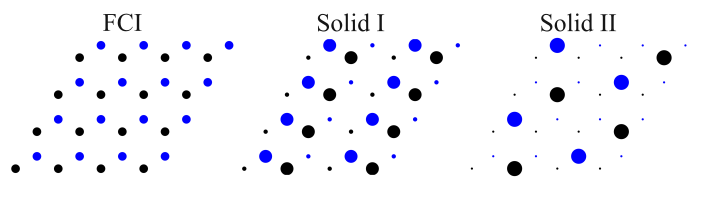}
        \caption{Real space occupation pattern of 3 phases, the black and blue color label A and B sublattice and the size of the circle represents the charge density}
        \label{fig:cdwpattern}
    \end{figure} 

 In our result, the interplay between nearest-neighbor(NN) ($V_1=4$) and next-nearest-neighbor(NNN) ($V_2$) repulsions stabilizes the Solid I and Solid II phases. The NNN interaction inherently favors a charge density wave (CDW) order with doubled periodicity relative to the primitive lattice vector. As $V_2$ increases, the approximate translational symmetry along the $r_1$ direction(under open boundary conditions) is initially broken, triggering the FCI-Solid I phase transition. Subsequently, the Solid I - Solid II transition occurs at higher $V_2$ values, where the charge occupation becomes fully polarized among the sublattice to minimize the energy penalty imposed by $V_2$, driving the system into the Solid II phase  The real-space density pattern of the Solid I and II is illustrated in Fig.\ref{fig:cdwpattern} \\

Fig.\ref{fig_S_phase_transition} (a) and (b) are the static charge order and density structure factor measured at $\mathbf{M_1}$ point. In FCI-Solid I phaes transition, the roton mode at $\mathbf{M_1}$ goes soft near phase boundary and gives rise to the charge order at $\mathbf{M_1}$ in Solid I phase, as observed from the formation of charge order (Fig.~\ref{fig_S_phase_transition} (a)) and the associated peak of density structure factor (Fig.~\ref{fig_S_phase_transition} (b)).

Similar feature of  roton-driven phase transition is observed in Solid I-Solid II phase transition, as shown in Fig.~\ref{fig_S_phase_transition} (c) and (d). The roton mode at $\mathbf{b_1}/4 + \mathbf{b_2}/2$ goes soft near phase boundary and give rise to the charge order at the same momentum in Solid II phase.

Thus we conclude that the transition points of FCI-Solid I and Solid I-Solid II phase transition are around $V_2 = 2.2$ and $V_2 = 3.3$ along the fixed $V_1 = 4$ path, respectively.

\subsection{Section V: Spectra along different paths in BZ}

\begin{figure}[ht!]
	\centering		
	\includegraphics[width=0.6\textwidth]{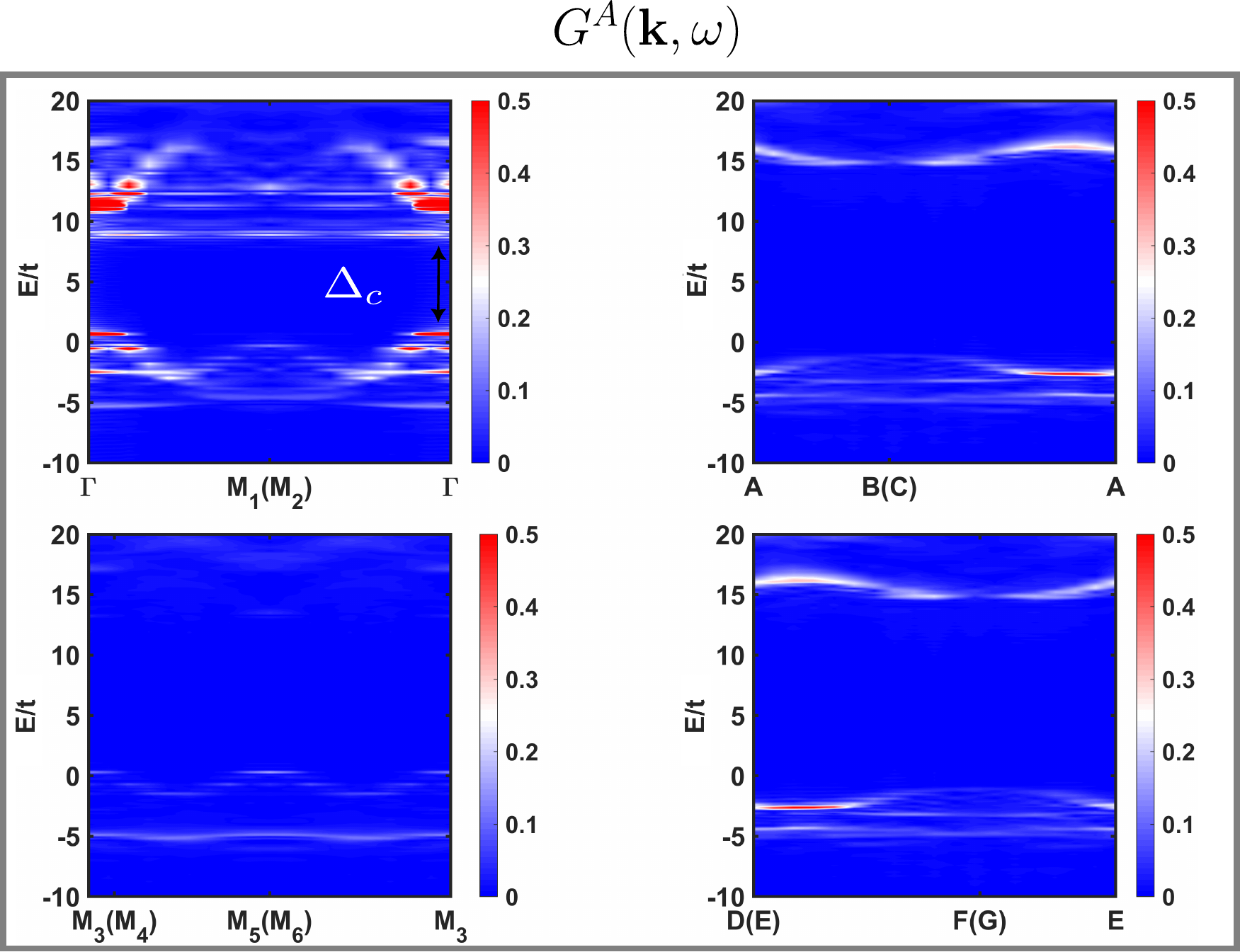}
	\caption{Single-particle spectrum inside the FCI phase ($V_1 = 4, V_2 = 1$) plotted along the 4 paths. The spectrum weight mainly concentrate at the minimal charge gap near $\Gamma$ point.The charge gap minimum at different paths is labeled with black arrows.}
    \label{fig_S_ehspectrum}
\end{figure}
In this section we show the spectra of the  single-particle excitation $G^A(\mathbf{k},\omega)$ along 4 different paths in the BZ in Fig.3 (d) of the main text.

The single particle spectrum $G^A(\mathbf{k},\omega)$ along the 4 paths are shown in Fig.~\ref{fig_S_ehspectrum}. The spectrum weights mainly concentrate at the minimal charge gap near $\Gamma$ point. We note that in our honeycomb lattice model, the band gap of the tight-binding Hamiltonian is around 2.5 which is larger than the roton gaps $\Delta_{n_1}$ and $\Delta_{n_2}$, while the charge gap is much larger than the band gap as shown in Fig.~\ref{fig_S_ehspectrum}.

\subsection{Section VI: Fermionic roton mode}
\begin{figure}[h!]
	\centering		
	\includegraphics[width=\textwidth]{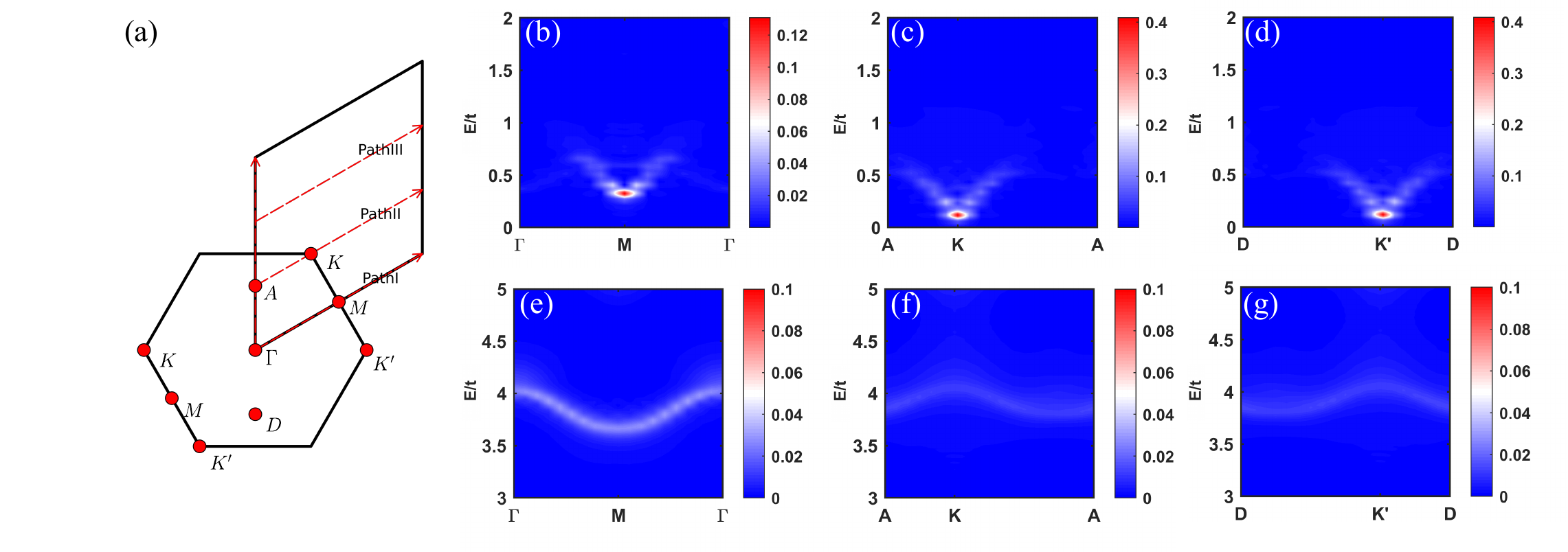}
	\caption{Magneto-roton for $1/3$ fermionic FCI at $V_1 = 4, V_2 = 1$. (a) (b), and (c) are the dynamic structure factor $S^A(\mathbf{k},\omega)$ along 3 different paths, respectively. (d) Illustrates 3 paths in BZ}
    \label{fig_S_fermionicroton}
\end{figure}	
In this section, we show the spectra of the magneto-roton of $1/3$ fermionic FCI. As shown in Fig.~\ref{fig_S_fermionicroton}.

Fig.~\ref{fig_S_fermionicroton}(b)(c)(d) display the $S^A(\mathbf{k},\omega)$ in energy range $\omega < 2$, which we attribute to intra-band dynamics. In the fermionic case, the magneto-roton's characteristic energy scale ($\Delta_r \sim 0.1$) is smaller than the bosonic case ($\Delta_r \sim 1$), although with the same interaction strength ($V_1 = 4, V_2 = 1$). The magnetoroton minimum is located at $K$ point and has higher energy at the $M$ point. Our magneto-roton dispersion qualitatively agrees with recent results based on the GMP ansatz with periodic potential~\cite{Kousa2025Theory}. 

Fig.~\ref{fig_S_fermionicroton}(b)(c)(d) display the $S^A(\mathbf{k},\omega)$ in higher energy scale. Such modes could arise from band mixing effect, which is beyond the GMP ansatz and hasn't been observed in our bosonic FCI simulations.

\subsection{Section VII: Evolution of roton gaps inside FCI phase}
\begin{figure}[ht!]
	\centering		
	\includegraphics[width=0.4\textwidth]{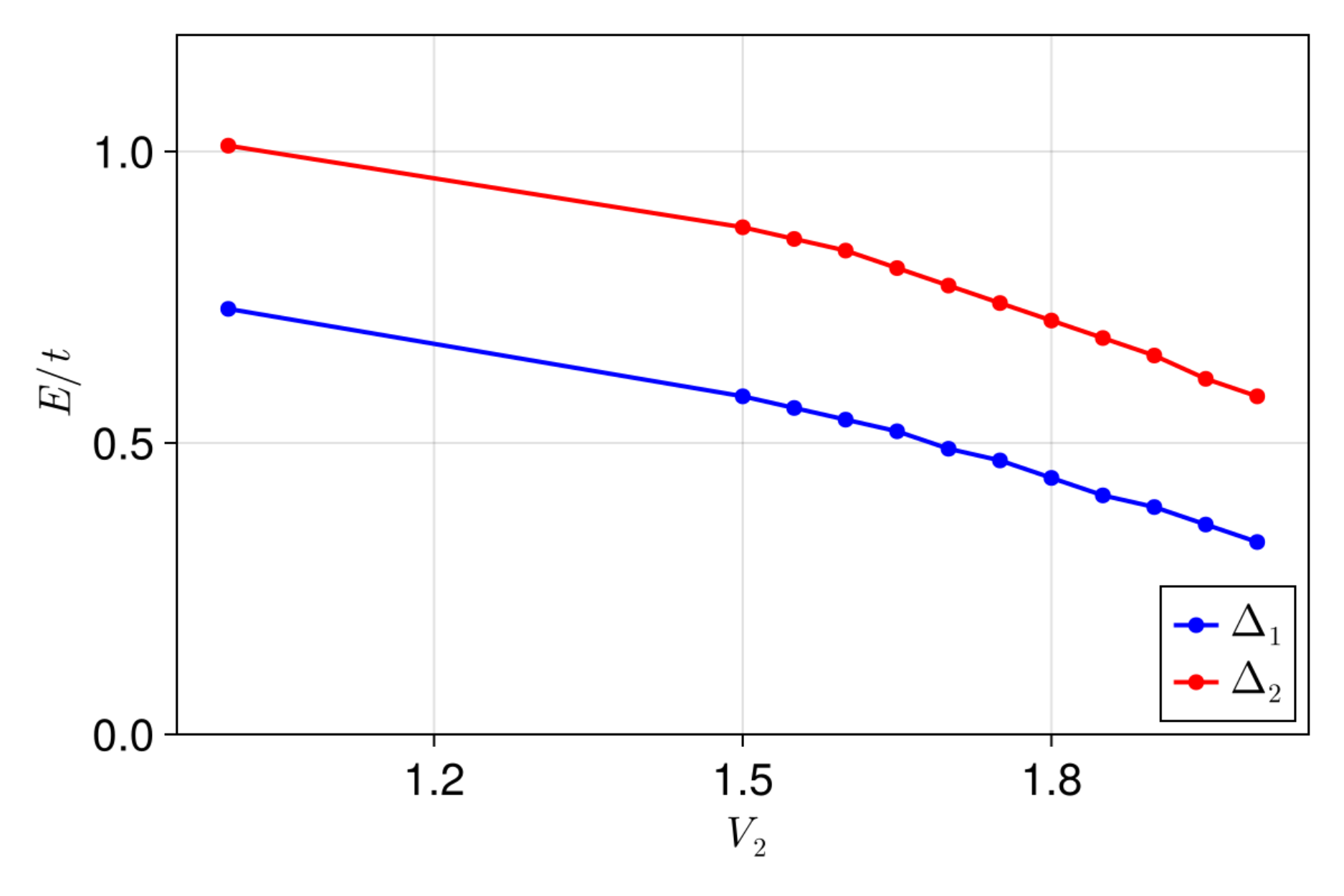}
	\caption{Evolution of the roton gaps in FCI phase. $\Delta_{n_1}$ and $\Delta_{n_2}$ represent the first and second roton gaps respectively. Their energies become lower while approaching the FCI-Solid I phase boundary.}
    \label{fig_S_rotongap_evolution}
\end{figure}	

In this section, we trace the evolution of the roton gaps $\Delta_{n_1}$ and $\Delta_{n_2}$ as the system approaches the phase boundary. As shown in Fig.~\ref{fig_S_rotongap_evolution}, the roton gaps $\Delta_{n_1}$ and $\Delta_{n_2}$ gradually decrease as the system approaches the phase boundary, which is consistent with the fact that the roton mode go soft near the phase boundary and give rise to the charge order in the CDW phase. $\Delta_{n_2}$ always remain larger than $\Delta_{n_1}$ and as discussed in the main text, it will trigger the transition from Solid I to Solid II after the condensation of $\Delta_{n_1}$.


\subsection{Section VIII: Benchmark with exact diagonalization(ED)}

To verify the accuracy of the TDVP calculation method, we use the ED Lanczos method to compute the local density of states within a $4\times 4\times 2$ cylindrical geometry for comparative analysis. The dynamic Green's function in this context can be represented by a continued fraction expansion within the tridiagonal basis of the Hamiltonian, employing the Lanczos iterative technique.  This approach allows for a straightforward calculation of real-frequency correlation functions. During our calculations, we set the Lorentz broadening factor to $\eta = 0.05t_1$. We Fourier transform the real-frequency correlation function to obtain the retard Greens function $G_{ii}(t) = \langle b_i^\dagger(t) b_i\rangle + \langle b_i b_i ^\dagger(t)\rangle$ where $i$ site locates at A sublattice of $(m=2,n=2)$th unitcell, the data is calculated at $V_1 = 4,V_2 = 2.7$, which is inside Solid I phase. 

Our findings demonstrate that the results derived from the TDVP method are quantitatively consistent with those obtained using the ED method.

\begin{figure}[h!]
	\centering		
	\includegraphics[width=0.4\textwidth]{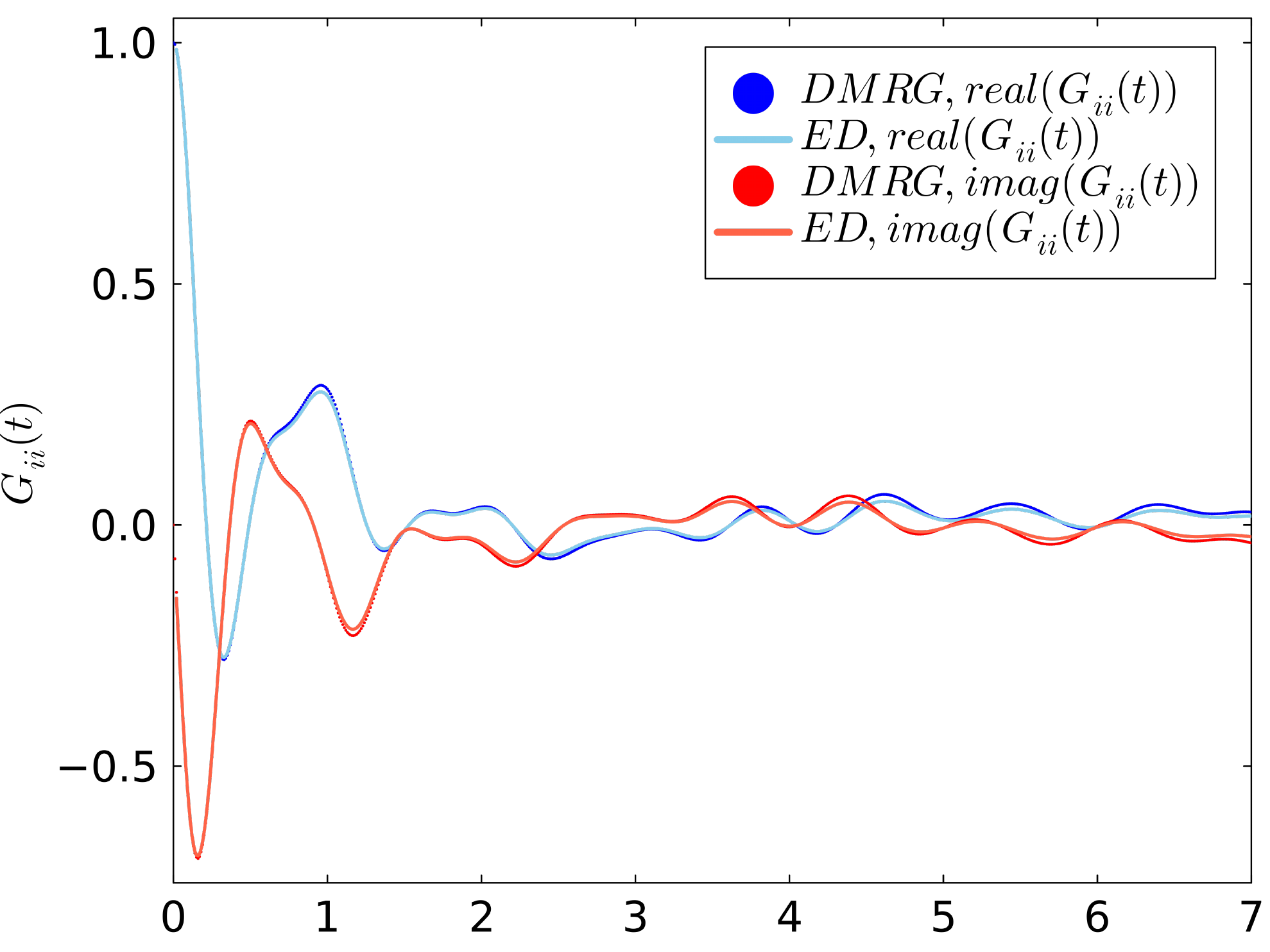}
	\caption{Time evolution of the real space correlation function $G_{ii}(t)$ for a $4\times 4\times 2$ cylindrical geometry. The results obtained using the TDVP method are consistent with those derived from the ED method.}
    \label{fig_S_rotongap_evolution}
\end{figure}

\end{widetext}

\end{document}